\documentclass[prb,twocolumn]{revtex4}
\usepackage{graphicx}
\usepackage{hyperref}
\usepackage{amssymb}
\usepackage{dcolumn}
\usepackage{float}
\usepackage{bm}
\usepackage{booktabs}
\usepackage{amsmath} 
\usepackage[utf8]{inputenc}
\usepackage{lmodern}
\usepackage{color} 
\definecolor{red}{rgb}{1.0,0.0,0.0}
\usepackage{makecell}
\usepackage{marginnote}

\DeclareMathAlphabet{\bi}{OML}{cmm}{b}{it}

\def\ba{\begin{aligned}}
\def\ea{\end{aligned}}
\def\be{\begin{equation}}
\def\ee{\end{equation}}
\def\bearr{\begin{eqnarray}}
\def\eearr{\end{eqnarray}}

\def\l{\left}
\def\r{\right}

\usepackage{tikz}
\begin{document}
\title{Berry curvature induced anisotropic magnetotransport in a quadratic triple-component fermionic system}
\bigskip
\author{Ojasvi Pal, Bashab Dey and Tarun Kanti Ghosh\\
\normalsize
Department of Physics, Indian Institute of Technology-Kanpur,
Kanpur-208 016, India}

\begin{abstract}
Triple-component fermions are pseudospin-1 quasiparticles hosted by certain three-band semimetals in the vicinity of their band-touching nodes [\href{https://doi.org/10.1103/PhysRevB.100.235201}{Phys. Rev. B {\bf 100}, 235201 (2019)}]. The excitations comprise of a flat band and two dispersive bands. The energies of the dispersive bands are  $E_{\pm}=\pm\sqrt{\alpha^2_n k^{2n}_\perp+v^2_z k^2_z}$ with $k_\perp=\sqrt{k^2_x+k^2_y}$ and $n=1,2,3$. In this work, we obtain the exact expression of Berry curvature, approximate form of density of states and Fermi energy as a function of carrier density for  any value of $n$. In particular, we study the Berry curvature induced electrical and thermal magnetotransport properties of quadratic $(n=2)$ triple-component fermions  using semiclassical Boltzmann transport formalism. Since the energy spectrum is anisotropic, we consider two orientations of magnetic field (${\bf B}$): (i) ${\bf B}$ applied in the $x$-$y$ plane and (ii)  ${\bf B}$ applied in the $x$-$z$ plane. For both the orientations, the longitudinal and planar magnetoelectric/magnetothermal conductivities 
show the usual quadratic-$B$ dependence and oscillatory behaviour with respect to the angle between the applied electric field/temperature gradient and magnetic field as observed in other topological semimetals. However, the out-of-plane magnetoconductivity  has an oscillatory dependence on angle between the applied fields for the second orientation but is angle-independent for the first one. 
We observe large differences in the magnitudes of transport coefficients for the two orientations at a given Fermi energy. 
A noteworthy feature of quadratic triple-component fermions which is typically absent in conventional systems is that certain transport coefficients and their ratios are independent of Fermi energy within the low-energy model.
\end{abstract}
\maketitle
\section{Introduction}
Topologically nontrivial gapless materials termed as topological semimetals have unearthed vast research interests in condensed matter physics within the past few  years. Some well-known examples are topological Dirac\cite{DSM1,DSM2,DSM3,DSM4,DSM5} and Weyl\cite{weyl1,weyl2,weyl3,weyl4,weyl5,weyl6,weyl7,weyl8} semimetals which can host emergent quasiparticle excitations analogous to the fermionic elementary particles in high-energy physics. They are characterized by symmetry-protected crossings of two energy bands at the Fermi level. In close vicinity of such band-touching points, the fermionic system can be effectively expressed in terms of pseudospin degrees of freedom, where the distinct eigenvalues of the pseudospin projection corresponds to different energy bands. 

Apart from the above mentioned half-integer spin Weyl and Dirac semimetals, recent studies\cite{Bradlyn} suggested the existence of another class of topological semimetals with three-, six-, or eight-fold degenerate points at the Fermi level that exhibit fermionic quasiparticles having no counterparts in high-energy physics. Furthermore, it was observed that various possible symmetries existing in symmorphic crystals, for instance, mirror and discrete rotational symmetries may lead to topologically protected three-fold degenerate crossing points \cite{symmetry,symmorphic}. It was predicted by first-principles calculations\cite{tcfs1,tcfs2,firstprinciple1,firstprinciple2} that several material candidates categorized under symmorphic space groups, such as  TaN\cite{tcfs2}, MoP\cite{eg-2}, WC\cite{eg-3}, RhSi\cite{eg-4}, RhGe\cite{eg-5} and ZrTe\cite{eg-6}  can host crossings of three bands in the neighborhood of the Fermi level. The low energy excitations in such systems behave like pseudospin-1 quasiparticles but obey fermionic statistics and are named as triple-component fermions\cite{bitan} (TCFs). The touching of these energy bands occurs only in pairs at specific points in the Brillouin zone whose location is decided by the underlying symmetry of the system. Such band-touching points are called triple-component points or nodes. 

The general low-energy Hamiltonian of such pseudospin-1 fermionic system in the vicinity of band-touching points can be effectively described as $H=d({\mathbf{k}}) \cdot  \mathbf{S}$, where  $d({\mathbf{k}})$ is a function of momenta and $\mathbf{S}$ are the three spin-1 matrices. 
The low-energy dispersion of the quasiparticle near each node appears as ${m}\sqrt{{\alpha}_{n}^{2}(k_{x}^{2}+k_{y}^{2})^{n}+{\nu}_{z}^{2}k_{z}^{2}}$ with $m=\pm1$ and $0$ representing the dispersive and flat bands respectively. Here, the different combination of crystal symmetries\cite{C4} impose a restriction on $n$, such that $n\leq3$. Triple-component fermions are predicted to occur in time-reversal symmetric materials such as Pd$_{3}$Bi$_{2}$S$_{2}$ and Ag$_{2}$Se$_{2}$Au with space group 199 and 214 respectively\cite{Bradlyn}. The experimental realization of materials having time-reversal symmetry (TRS) broken TCFs is still missing.

The band touching nodes of topological semimetals are sources and drains of Berry curvature \cite{BC,xiao} which acts as a fictitious magnetic field in the momentum space and leads to several intriguing transport phenomena such as anomalous Hall effect (AHE)\cite{xiao,AHE1,AHE2,AHE3,AHE4}, anomalous thermal  Hall effect (ATHE)\cite{ATHE1,ATHE2,ATHE3}, negative longitudinal magnetoresistance (MR)\cite{1,sasaki,2,3,4,5,6,7,8,lu,pal}, planar Hall effect (PHE)\cite{PHE1,PHE2,PHE3,PHE4,PHE5} and  planar thermal Hall effect (PTHE)\cite{PTHE1,PTHE2,PTHE3}. The Berry curvature induced magnetotransport phenomena in type-I multi-Weyl semimetals has been studied recently\cite{mutiweyl}. 
The triple component nodes also give rise to Berry curvature in the two dispersive bands and affect the transport phenomena in presence of magnetic field. Each node hosts a monopole charge of $|2n|$. The magnetotransport properties\cite{bitan} induced by Berry curvature, ballistic transport across junction\cite{diptiman} and plasmons\cite{plasmontcf} in TRS-broken triple-component fermionic system for $n=1$ case has been studied recently. In this work, we consider the low-energy model of TRS-broken TCFs specifically for $n=2$ case (quadratic TCFs) and study its Berry curvature induced electrical and thermal magnetotransport properties. The dispersion of quadratic TCFs has different scaling with momenta along different directions which allows us to use two orientations of planar Hall geometry to develop our understanding towards the behavior of magnetotransport coefficients in response to linear dispersion along one symmetry direction and quadratic dispersion for the other two directions. Hence measurement of transport coefficients in presence of small magnetic field may act as a probe for detecting different symmetry axes of the system. 

This paper is organized as follows: In Sec. \ref{II}, we present a discussion on the low-energy model for generalized triple component fermions. In Sec. \ref{III}, we provide a review of the semiclassical Boltzmann transport formalism incorporating the Berry curvature effects. In Sec. \ref{IV}, we obtain the expressions of  magnetoconductivities in the presence of in-plane electric and magnetic fields and further discuss the magnetic field and angular dependence of our numerically computed results of  longitudinal
magnetoconductivity (LMC) and planar Hall conductivity (PHC) for two different orientations of electric and magnetic fields. Section \ref{V} is dedicated to the detailed discussion of magnetoresistance. The results of positive longitudinal
magneto-thermal conductivity (LMTC)  and planar thermal Hall conductivity (PTHC) are presented in Sec. \ref{VI}. In Sec. \ref{comparesec}, we make a quantitative comparison of the transport coefficients calculated for the two configurations. Finally, we summarize our main results in Sec. \ref{VII}.
\section{Low-energy model for generalized triple component fermions}\label{II}
The effective low-energy Hamiltonian for TRS broken generalized TCFs can be expressed as\cite{bitan}
\begin{equation}\label{hamiltonian1}
H_{n,\tau}({\mathbf{k}}) = d_{1}^{n}({\mathbf{k}}){S_x}+ d_{2}^{n}({\mathbf{k}}){S_y}+ {\tau} d_{3}({\mathbf{k}}){S_z},
\end{equation}
where  $\tau=\pm1$ is the valley index representing the two valleys or triple-component points and $S_x$, $S_y$, and $S_z$ are the components of spin-1 matrix operator given by
\begin{equation}
\begin{aligned}
S_x &= \frac{1}{\sqrt{2}}
 \left (\begin{matrix}
  {0}&{1}&{0}  \\
{1} &{0}&{1}\\
{0} &{1}&{0}\\
\end{matrix} \right),\hspace{0.5cm} S_y &= \frac{1}{\sqrt{2}}
 \left (\begin{matrix}
  {0}&{-i}&{0}  \\
{i} &{0}&{-i}\\
{0} &{i}&{0}\\
\end{matrix} \right),\\
&S_z = \left (\begin{matrix}
  {1}&{0}&{0}  \\
{0} &{0}&{0}\\
{0} &{0}&{-1}\\
\end{matrix} \right).
\end{aligned}
\end{equation}
Here, the different combinations of symmetries in pseudospin-1 fermionic systems impose a restriction on $n$, i.e., $n\leq3$. The coefficients $d_{1,2,3}^{n}({\mathbf{k}})$ are functions of momenta which vary according to the given choice of $n$\cite{bitan}. The Hamiltonian in Eq.(\ref{hamiltonian1}) can be rewritten compactly as\begin{equation}\label{polar}
H_{n,\tau}({\mathbf{k}}) = \alpha_{n}{k}_{\perp}^{n}({S_x}\cos{n}\phi+{S_y}\sin{n}\phi) + {\tau} \nu_{z}k_{z}{S_z},
\end{equation}
where $k_{\perp}=\sqrt{k_{x}^{2}+k_{y}^{2}}$, ${\nu_z}/\hbar$ and ${\alpha_1}/\hbar$ have the dimension of velocity, ${\alpha_2}/\hbar^{2}$ has the dimension of inverse mass  and $\alpha_3$ has the dimension of inverse of density of states. The diagonalization of the Hamiltonian leads to the energy spectra in the neighborhood of each valley as
\begin{equation}
\epsilon_{n}^{m}(\mathbf{k})={m}\sqrt{{\alpha}_{n}^{2}k^{2n}_\perp+{\nu}_{z}^{2}k_{z}^{2}}, 
\end{equation}
where ${m}= -1,0,1$ is the pseudospin projection.  
The dispersive bands with ${m}= \pm1$ indicate the conduction and valence bands respectively and ${m}=0$ represents the flat band at zero energy. For $n=1$, the two dispersive bands scale linearly with all the three components of momentum. These excitations are called linear TCFs. For $n=2$, the excitations are named as quadratic TCFs since the energy scales linearly with $k_z$ but quadratically with $k_\perp$.  Cubic TCFs refer to the $n=3$ case, where the energy scales linearly with $k_z$ but cubically with $k_\perp$.

The eigenstates corresponding to the three bands can be written as
\begin{equation}
|\psi_{n,\tau,\bf k}^+\rangle = 
 \left (\begin{matrix}
  {(\frac{1+\tau\cos{\gamma_{n}}}{2} ) } \\
{ \frac{\sin{\gamma_{n}}}{\sqrt{2}}}e^{in\phi}\\
 {(\frac{1-\tau\cos{\gamma_{n}}}{2}) e^{2in\phi}}
\end{matrix} \right),
\end{equation}
\begin{equation}
|\psi_{n,\tau,\bf k}^-\rangle = 
 \left (\begin{matrix}
  {(\frac{1-\tau\cos{\gamma_{n}}}{2} )} \\
{- \frac{\sin{\gamma_{n}}}{\sqrt{2}}}e^{in\phi}\\
 {(\frac{1+\tau\cos{\gamma_{n}}}{2}) e^{2in\phi}}
\end{matrix} \right),
\end{equation}
and\\
\begin{equation}
|\psi_{n,\tau,\bf k}^{0}\rangle = 
 \left (\begin{matrix}
  {-\frac{\sin{\gamma_{n}}}{\sqrt{2}}} \\
  {\tau\cos\gamma_{n}}{e^{in\phi}}\\
 { \frac{\sin{\gamma_{n}}}{\sqrt{2}}e^{2in\phi}}
\end{matrix} \right).\\
\end{equation}
Here, we define $\gamma_{n}={\tan}^{-1}[({\alpha_n}/{\nu_z})\tan\theta(k\sin\theta)^{n-1}]$. The expectation value of components of the spin-1 matrix operator with respect to the eigenstates $|\psi_{n,\tau,\bf k}^m\rangle$ are calculated as: ${\langle{S_x}\rangle}_{m}=m(\sin{\gamma_{n}})\cos{n}\phi$, ${\langle{S_y}\rangle}_{m}=m(\sin{\gamma_{n}})\sin{n}\phi$ and ${\langle{S_z}\rangle}_{m}=m\cos{\gamma_{n}}$.

The touching points of non-degenerate pair of bands in the momentum space act as a source and sink of the Abelian Berry curvature. The Berry curvature of the $m$th band for the Hamiltonian given by Eq. (\ref{polar}) can be calculated using $\bm{\Omega}_{n,\tau,\mathbf{k}}^{m} = i{ \boldsymbol{\nabla}}_{\mathbf{k}}\times \langle \psi^{m}_{n,\tau,\mathbf{k}}|  \boldsymbol{\nabla}_\mathbf{k} |\psi^{m}_{n,\tau,\mathbf{k}}\rangle $ as\cite{xiao} 
\begin{widetext}
\begin{equation}
\bm{\Omega}_{n,\tau,\mathbf{k}}^{m} = m n\tau{\nu_z}{\alpha}_{n}^{2}{k}^{(2n-4)}\left[\frac{-(\sin{\theta})^{2n-2}[1+(n-1)\cos^2{\theta}]\hat{{k}}+[(n-1)\cos\theta{(\sin\theta)}^{2n-1}]\hat{{\theta}}}{{\left({{\alpha_{n}^{2}}k^{2n-2}(\sin{\theta})^{2n}+\nu_{z}^{2}\cos^2{\theta}}\right)}^{3/2}}
\right].
\end{equation}
\end{widetext}

\begin{figure}[htbp]
\includegraphics[trim={0cm 0cm 0cm 0cm},clip,width=8.5
cm]{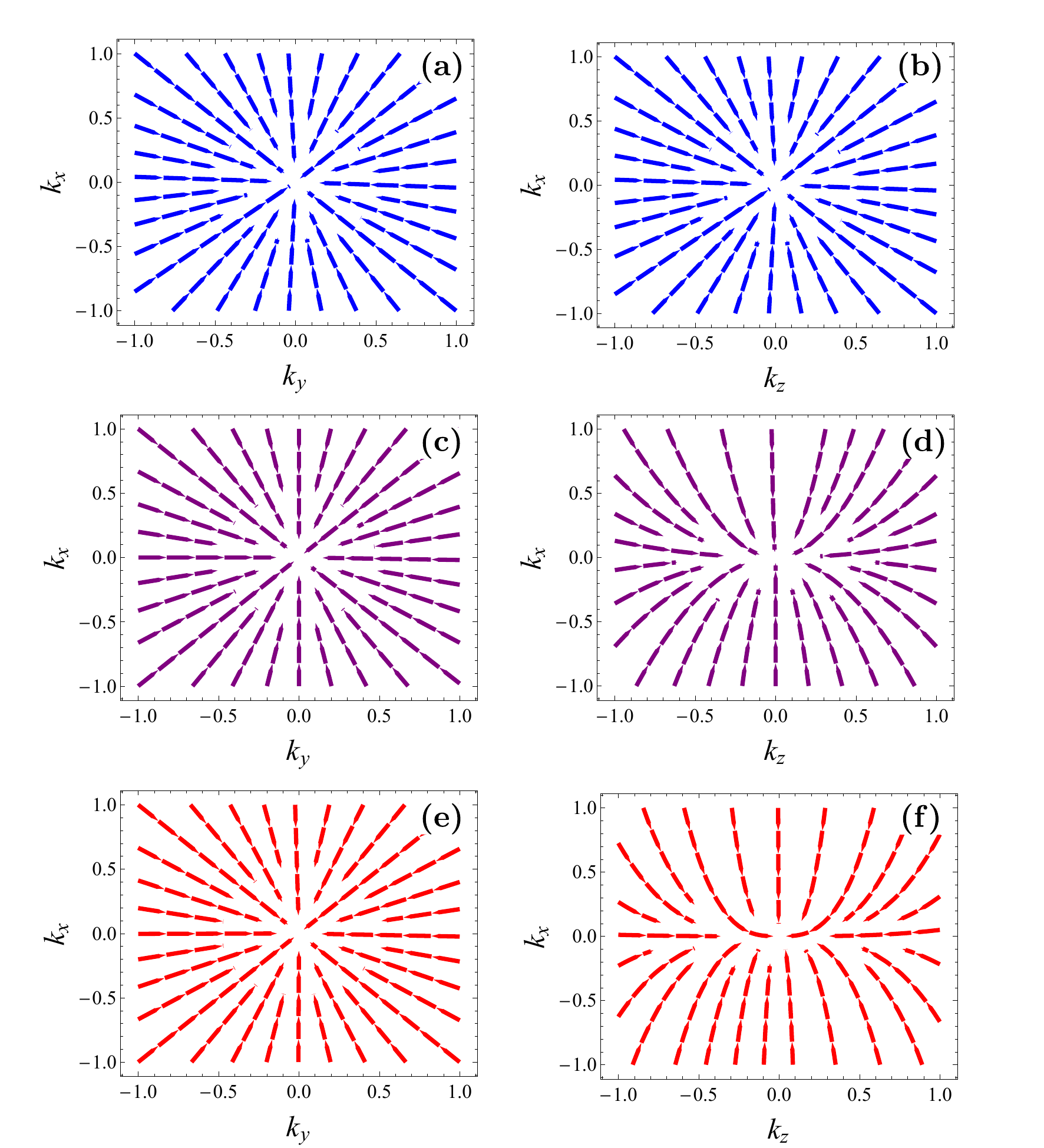}
\caption{Vector plots of the Berry curvature of the conduction band for triple-component semimetals -- upper row: $n=1$, middle row: $n=2$ and lower row: $n=3$ in $k_x$-$k_y$ plane and $k_x$-$k_z$ plane respectively.} 
\label{0000}
\end{figure}

The Berry curvature of the flat band is identically zero for all momenta, but is finite for the two dispersive bands carrying opposite signs. The Berry curvature of linear TCFs contains only the radial component while that of quadratic and cubic TCFs have angular components apart from the radial ones. In the $k_x$-$k_y$ plane, the Berry curvature varies as $\sim m n \tau \nu_z  /(\alpha _n k^{n+1})\hat{k}$. 
Integrating the Berry curvature over a unit sphere enclosing the triple-component point defines the monopole charge, given by
\begin{equation}
\aleph=\frac{1}{2\pi}\int_{A} \bm{\Omega}_{\mathbf{k}} \cdot d\mathbf{A}=2n.
\end{equation}
The vector plots of  Berry curvature of conduction band for TCFs with $n=1,2,3$ and $\tau=1$ in $k_x$-$k_y$ plane and $k_x$-$k_z$ plane are shown in Fig. \ref{0000}. The angular dependence of Berry curvature for $n=2$ and $n=3$ can be seen in the plots for $k_x$-$k_z$ plane. The converging nature of arrows represents the sink of Berry curvature at the origin. 

The density of states (DOS) for linear TCFs including valley degeneracy is given by
\begin{equation}
D_{1}({\mu})=D_{1}'\frac{{{\tilde{\mu}}^2}}{2\pi^2},
\end{equation}
where  $D_{1}'=2k'^{2}{\nu}_{z}/{\alpha}_{1}^{2}$ and $\tilde{{\mu}}=\mu/{\mu}'_{1}$ with ${\mu}'_{1}=k'{\nu}_{z}$ being an energy scale and $k'$ an arbitrary wavenumber scale. The DOS for quadratic TCFs can be obtained as 
\begin{equation}{\label{dos}}
\begin{aligned}
D_{2}({\mu})=D_{2}'\int_{0}^{\pi} d\theta
\left(\frac{\tilde{\mu}\sqrt{-\cos^2{\theta}+\sqrt{\cos^4{\theta}+4\tilde{\mu}^{2}\sin^4\theta}}}{4\sqrt{2}{{\pi}^2}\sin{\theta}\sqrt{\cos^4{\theta}+4\tilde{\mu}^{2}\sin^4\theta}}\right)
\end{aligned}
\end{equation}
\begin{figure}[htbp]
\includegraphics[trim={2cm 4cm 2cm  0.5cm},clip,width=9
cm]{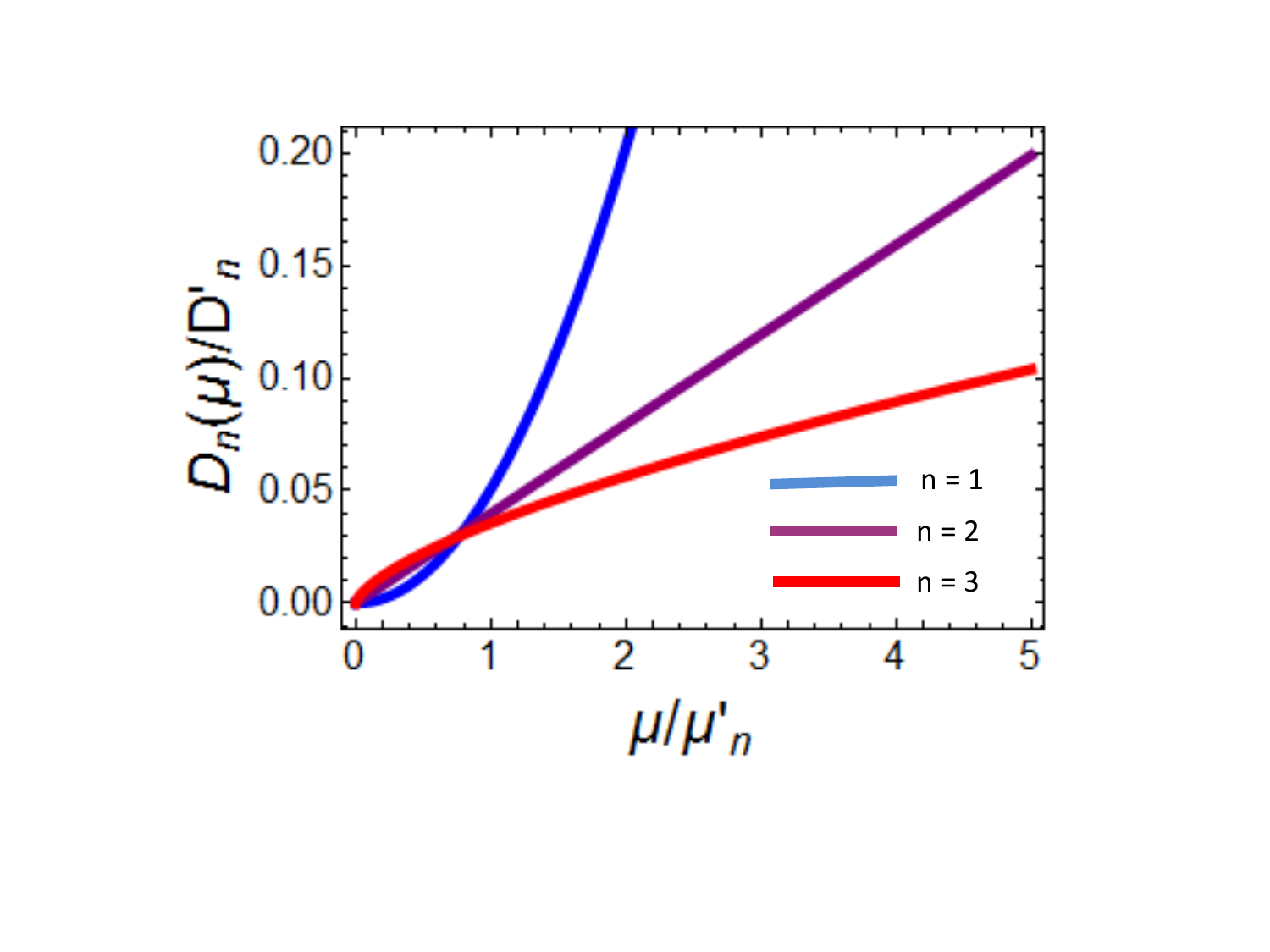}
\caption{Density of states as a function of the Fermi energy for triple-component semimetals with $n=1,2,3$.}
\label{00}
\end{figure}
\begin{figure}[htbp]
\includegraphics[trim={2cm 3.3cm 2cm  0.5cm},clip,width=9
cm]{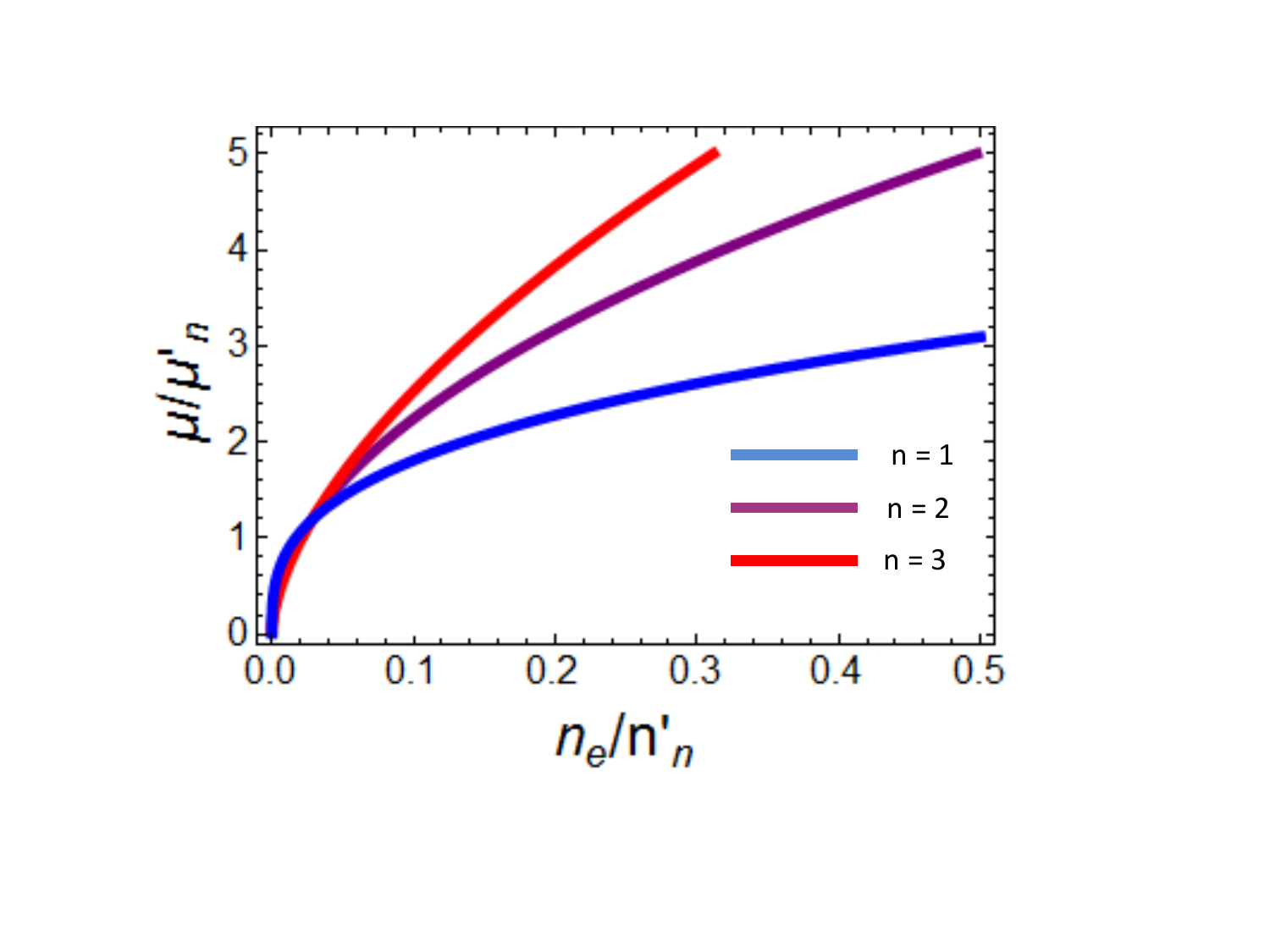}
\caption{ Plots of variation of the Fermi energy with electron density in the $T\rightarrow0$ limit for  linear, quadratic and cubic TCFs.}
\label{000}
\end{figure}
Here, we have defined  $D_{2}'=2{\nu}_{z}/{\alpha}_{2}^{2}$ and $\tilde{{\mu}}=\mu/{\mu}'_{2}$ with ${\mu}'_{2}={\nu}_{z}^{2}/{\alpha}_{2}$. Unlike Weyl semimetals and 3D free electron gas, the variation of DOS in quadratic TCFs is approximately linear in energy. For cubic  TCFs, the integral expression of DOS is cumbersome and hence not presented here.  We numerically find that $D_{3}({\mu})/{D_{3}'}$ varies approximately as $\tilde{{\mu}}^{2/3}$, where  $D_{3}'=2/{\alpha}_{3}$  and $\tilde{{\mu}}=\mu/{\mu}'_{3}$ with ${\mu}'_{3}={\nu}_{z}^{3/2}/{\alpha}_{3}^{1/2}$. Hence, the DOS of generalized TCFs varies with Fermi energy as $D_{n}(\mu)\sim {\mu}^{2/n}$, where $n=1, 2, 3$, as shown in Fig. \ref{00}. The electron density $n_{e}$ at temperature $T$ can be obtained from the following condition,
\begin{equation}
n_{e}=\int_{0}^{\infty} f_{eq}(\mu)D({\mu})d\mu
\end{equation}
where $f_{eq}(\mu)$ is equilibrium Fermi-Dirac distribution function. In linear TCFs, the Fermi energy varies with electron density at $T\rightarrow0$ as $\mu/{\mu}'_{1}\sim ({{n_e}/{n'_{1}}})^{1/3}$ with ${n'_{1}}=k'^{3}{\nu}_{z}^{2}/{\alpha}_{1}^{2}$. The quadratic and cubic  TCFs  show the variation with electron density as $\mu/{\mu}'_{2}\sim ({{n_e}/{n'_{2}}})^{1/2}$ and $\mu/{\mu}'_{3}\sim ({{n_e}/{n'_{3}}})^{3/5}$ with  ${n'_{2}}={({\nu}_{z}/{\alpha_{2}})}^{3}$ and ${n'_{3}}={({\nu}_{z}/{\alpha_{3}})}^{3/2}$ respectively. 
In Fig. \ref{000}, we have plotted  the Fermi energy as a function of electron density for generalized TCFs which scales as $\mu\sim({n_{e}})^{\frac{n}{n+2}}$.

\section{Theoretical formulation}\label{III}
We will use Boltzmann transport formalism to obtain the magnetotransport properties of quadratic TCFs. The semiclassical Boltzmann transport approach is valid for weak magnetic fields such that $\omega_{c}\tau\ll1$, where $\omega_{c}$ is the cyclotron frequency and $\tau$ is the average time between two successive collisions. In this limit, the Landau quantization gets wiped out by disorder effects.

The phenomenological Boltzmann transport equation (BTE) for the non-equilibrium distribution function ${f}_{\mathbf{r},\mathbf{k},t}^{m}$ is given by\cite{ashcroft}
 \begin{equation}
\left(\frac{\partial }{\partial t}+\mathbf{\dot r}^{m} \cdot { \boldsymbol{\nabla}}_{\mathbf{r}} + \mathbf{\dot k}^{m} \cdot{ \boldsymbol{\nabla}}_{\mathbf{k}}\right ){f}_{\mathbf{r},\mathbf{k},t}^{m}= {I}_{\textnormal{coll}} \{{{f}_{\mathbf{r},\mathbf{k},t}^{m}}\}, 
\end{equation}
where right-hand side represents the collision integral and $m$ is the band index. Under the relaxation-time approximation, the collision integral takes the form, ${I}_{\textnormal{coll}} \{{{f}_{\mathbf{k}}^{m}}\}= -\frac{{f}_{\mathbf{k}}^{m}-{f}_{\textnormal{eq}}^{m}}{{\tau}_{\mathbf{k}}^{m}}$, where the scattering time scale  ${{\tau}_{\mathbf{k}}^{m}}$ is the  intranode relaxation time and ${f}_{\textnormal{eq}}^{m}={[1+ e^{\beta({\epsilon}_{\mathbf{k}}^{m}-\mu)}]}^{-1}$ is the equilibrium Fermi-Dirac distribution function. We will ignore the momentum dependence of relaxation time and assumed it to be constant throughout the paper. In the steady-state condition, the BTE takes the following form
\begin{equation}{\label{BTE}}
(\mathbf{\dot r}^{m} \cdot\nabla_{\mathbf{r}} + \mathbf{\dot k}^{m} \cdot\nabla_{\mathbf{k}}){f}_{{\mathbf{k}}}^{m} =-\frac{{f}_{{\mathbf{k}}}^{m}-{f}_{\textnormal{eq}}^{m}}{\tau}.
 \end{equation}
The modified semiclassical equations of motion for an electron incorporating the Berry curvature effects are given by\cite{xiao,EOM,moore}
\begin{equation}{\label{EOM1}}
\mathbf{\dot r}^{m}=\frac{1}{{D}_{\mathbf{k}}^{m}}\bigg[{\mathbf{v}}_{\mathbf{k}}^{m}+\frac{e}{\hbar}(\mathbf{E} \times \bm{\Omega}_{\mathbf{k}}^{m})+ \frac{e}{\hbar}({\mathbf{v}}_{\mathbf{k}}^{m} \cdot \boldsymbol{\Omega}_{\mathbf{k}}^{m})\mathbf{B}\bigg],
\end{equation}
\begin{equation}{\label{EOM2}}
\hbar \mathbf{\dot k}^{m} = \frac{1}{{D}_{\mathbf{k}}^{m}}\bigg[-e \mathbf{E}- e({\mathbf{v}}_{\mathbf{k}}^{m} \times \mathbf{B})- \frac{e^2}{\hbar}(\mathbf{E} \cdot \mathbf{B}) \boldsymbol{\Omega}_{\mathbf{k}}^{m}\bigg],
\end{equation}
where ${D}_{\mathbf{k}}^{m} = ({1+\frac{e\mathbf{B} \cdot \bm{\Omega}_{\mathbf{k} }^{m}}{\hbar}})$ is the phase space factor which alters the invariant phase space volume for Berry curvature modified dynamics as $[d\mathbf{k}]={D}_{\mathbf{k}}^{-1}[d\mathbf{k}]$\cite{mdos}, where $[d\mathbf{k}]={d^3}k/{(2\pi)}^3$. To balance this changed phase space volume, the density of states is multiplied by ${D}_{\mathbf{k}}$, so that the number of states in the volume element remains unchanged.

Here, the second term of Eq. (\ref{EOM1}) is the anomalous Hall velocity\cite{xiao,AHE1,AHE2,AHE3} arising due to Berry curvature and the third term $({\mathbf{v}}_{\mathbf{k}}^{m} \cdot \boldsymbol{\Omega}_{\mathbf{k}}^{m})\mathbf{B}$ give rise to the chiral magnetic effect\cite{CME1,CME2}. In Eq. (\ref{EOM2}), the first two terms describes the Lorentz force whereas the last term is responsible for chiral anomaly effect\cite{CME2,chiral1,chiral2,chiral3}. 

Substituting Eqs. (\ref{EOM1}) and (\ref{EOM2}) in Eq. (\ref{BTE}), we obtained the following expression of non-equilibrium distribution function defined upto linear order in $\mathbf{E}$ and ${\nabla}{T}$ as
\begin{equation}{\label{NDF}}
\begin{aligned}
{f}_{{\mathbf{k}}}^{m}=&{{f}}_{\textnormal{eq}}^{m}+\bigg[\frac{\tau}{{D}_{\mathbf{k}}^{m}}
\left(-e \mathbf{E}-\frac{({\epsilon}_{\mathbf{k}}^{m}-\mu)}{T}\boldsymbol{\nabla}{T}\right)\\
&\cdot \bigg({\mathbf{v}}_{\mathbf{k}}^{m}+\frac{e}{\hbar}{\mathbf{B}}
({{\mathbf{v}}_{\mathbf{k}}^{m}}\cdot \bm{\Omega}_{\mathbf{k}}^{m}) \bigg)+ {\mathbf{v}}_{\mathbf{k}}^{m}\cdot \mathbf{\Gamma}_{\mathbf{k}}^{m} \bigg]\bigg(-\frac{{\partial{{f}_{\textnormal{eq}}^{m}}}}{\partial{{{\epsilon}}_{\mathbf{k}}^{m}}} \bigg).
\end{aligned}
\end{equation}
The second term in the above expression denotes the deviation from equilibrium.  The term $({\mathbf{v}}_{\mathbf{k}}^{m}\cdot \mathbf{\Gamma}^{m})$ represents the Lorentz force contribution whose analytical expressions are given in the Appendix \ref{A}\cite{PHE1,PTHE3}. Since the prime focus of this paper is to study the effect of Berry curvature in transport coefficients, we have not considered the Lorentz force contribution to the conductivity.
The linear response equations for the charge current ($\mathbf{j}^{e}$) and thermal current ($\mathbf{j}^{q}$) to the external fields can be written as\cite{ashcroft}
\begin{equation}\label{j_e}
 {j}_{a}^{e}=L_{ab}^{11}{E_b}+L_{ab}^{12}(-\nabla_{b}T),
 \end{equation}
 and
 \begin{equation}\label{j_q}
 {j}_{a}^{q}=L_{ab}^{21}{E_b}+L_{ab}^{22}(-\nabla_{b}T),
 \end{equation}
where $a$ and $b$ are the spatial indices running over $x$, $y$ and $z$. The set of $L$ represents the transport coefficients, for instance, $L^{11}$ and  $L^{22}$ are the electrical and thermal conductivities respectively. The thermopower is defined as $S={L^{12}}/{L^{11}}$. The Onsager's relation relates the $L^{12}$ and $L^{21}$ as $L^{21}=TL^{12}$.

In the subsequent sections, we study the Berry curvature driven Boltzmann transport phenomena for quadratic TCFs.
\section{electrical transport}\label{IV}
In the absence of thermal gradient, the charge current is defined as
\begin{equation}{\label{j_electric}}
{\mathbf{j}} = -e\sum_{m}\int[d\mathbf{k}]{D}_{\mathbf{k}}^{m}({\dot{{\mathbf{r}}}^{m}}){f}_{{\mathbf{k}}}^{m},
\end{equation}
Substituting Eqs. (\ref{EOM1}) and (\ref{NDF}) in Eq. (\ref{j_electric}), we obtain the general expression of Berry curvature dependent electric conductivity as:
\begin{equation}{\label{sigma_ij}}
\begin{centering}
\begin{aligned}
\sigma_{ij} & = -\frac{e^{2}}{\hbar}\sum_{m}\int[d\mathbf{k}]\epsilon_{ijl}(\Omega_{{\mathbf{k}}}^{{m},l}){f}_{\textnormal{eq}}^{m}\\
&+e^{2}\tau\sum_{m}\int[d\mathbf{k}]{({D}_{\mathbf{k}}^{m})}^{-1}\bigg({{v}}_{i}^{m}+\frac{e}{\hbar}{B_{i}}
({{\mathbf{v}}_{\mathbf{k}}^{m}}\cdot \bm{\Omega_\mathbf{k}^{m}})\bigg)\\
& \bigg({{v}}_{j}^{m}+\frac{e}{\hbar}{B_{j}}
({{\mathbf{v}}_{\mathbf{k}}^{m}}\cdot \bm{\Omega_\mathbf{k}^{m}})\bigg)\bigg(-\frac{{\partial{{f}_{\textnormal{eq}}^{m}}}}{\partial{{\epsilon}_{\mathbf{k}}^{m}}} \bigg),
\end{aligned}
\end{centering}
\end{equation}
where $\epsilon_{ijl}$ is the Levi-Civita tensor and $\Omega_{{\mathbf{k}}}^{{m},l}$ denotes the $l$-component of Berry curvature. We expand the term $({D}_{\mathbf{k}}^{m})^{-1}$ upto quadratic order in ${\bf B}$-field as $(1+({e}/{\hbar}){\mathbf{B} \cdot \bm{\Omega}_{\mathbf{k} }^{m}})^{-1}=1-({e}/{\hbar}){\mathbf{B} \cdot \bm{\Omega}_{\mathbf{k} }^{m}}+({e^2}/{{\hbar}^{2}})({\mathbf{B} \cdot \bm{\Omega}_{\mathbf{k}}^{m}})^{2}$ in order to express the conductivity in power of magnetic field by separating various $B$-contributions. For the above expansion to converge, we must have $\left(\frac{e{B}{\Omega_{z}}}{\hbar }\right) << 1$. We find that the above condition is satisfied when  $\tilde{\mu}>>3.5$  for $\tilde{B}=2$, where $\tilde{B}=B/B_{0}$ and $\tilde{\mu}=\mu/\mu_{0}$ with $B_{0}=\nu_{z}^{2}\hbar/e\alpha_{2}^{2}$  and $\mu_{0}=\nu_{z}^{2}/\alpha_{2}$ being the magnetic field and energy scales respectively, of the given system. Therefore, for the rest of our calculations, we choose $\tilde{B}=2$ and $\tilde{\mu}=10$.

We emphasize that out of three bands, only the two dispersive bands participate in the  transport properties. In this work, we choose $\mu>0$ and thus Eq. (\ref{sigma_ij}) will have contribution only from conduction band $(m=1)$. Therefore, we drop the band index $m$ for rest of the paper to investigate these quantities. The first term of Eq. (\ref{sigma_ij}) represents the Berry curvature induced  intrinsic anomalous Hall conductivity. Since we are interested in calculation of transport coefficients in presence of magnetic field, we neglect the anomalous term. 

\begin{figure}[htbp]
\includegraphics[trim={0cm 9.6cm 0cm  0cm},clip,width=8.5
cm]{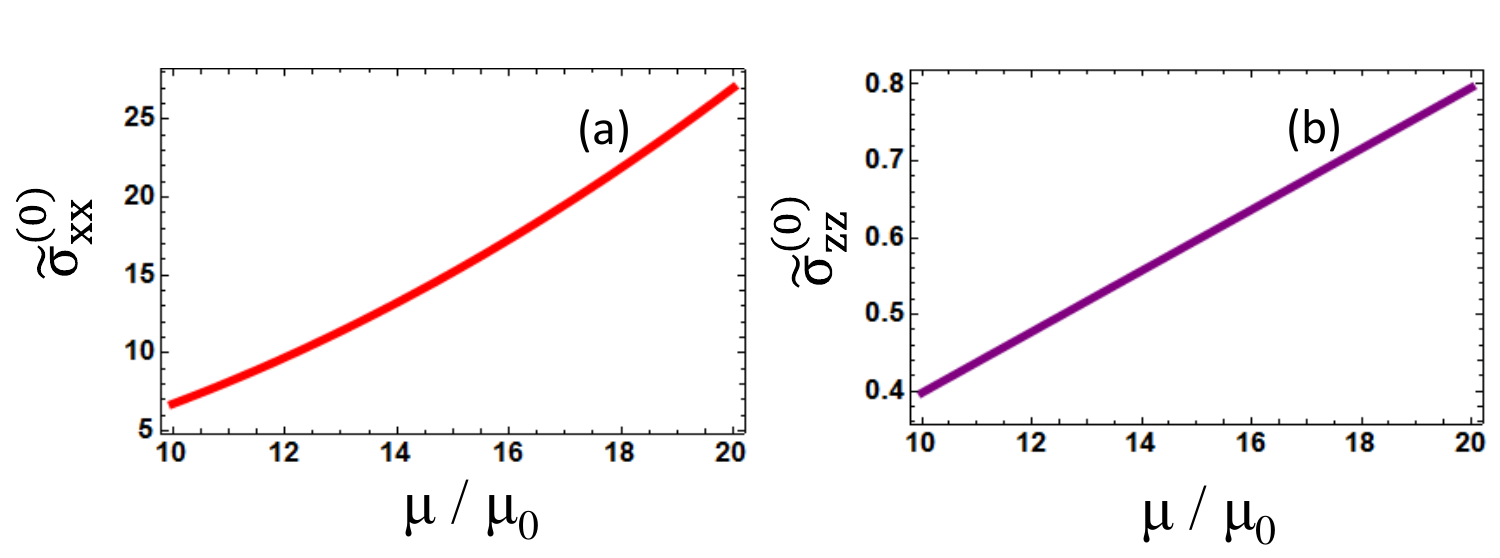}
\caption{(a)-(b) The variation of Drude conductivities with the Fermi energy for the quadratic TCFs, ${\sigma}_{xx}^{(0)}$ shows more increase with the Fermi energy as compared to ${\sigma}_{zz}^{(0)}$. We have defined $\tilde{\sigma}=\sigma/\sigma_{0}$ with $\sigma_{0}=\tau{e^2}{\nu_{z}^{3}}/{{\hbar^{2}}{\alpha_{2}^{2}}}$.} 
\label{003}
\end{figure}

It is to be noted that the magnetotransport results will be  qualitatively same for both low-energy model and lattice model as long as Fermi energy is close to the nodes. 
However, the calculation of anomalous Hall conductivity using low-energy model  gives completely different result as that obtained from the  tight-binding model.  In low-energy model, anomalous Hall conductivity trivially vanishes, whereas for lattice model it is finite and proportional to the distance between the nodes\cite{bitan}. 
\begin{center}
\textbf{Drude conductivities}
\end{center}
In the absence of magnetic field, the diagonal components of conductivity matrix are called Drude conductivity ($\sigma_{ii}^{(0)}$). We compute the Drude conductivities of quadratic TCFs in this section and discuss their dependence on Fermi energy. In the wake of anisotropic energy spectrum of quadratic TCFs, we have $\sigma_{xx}^{(0)}=\sigma_{yy}^{(0)}\neq\sigma_{zz}^{(0)}$. The general form of Drude conductivity is given by
\begin{equation}{\label{sigma_ij0}}
{\sigma}_{ii}^{(0)}= e^{2}\tau\int\frac{d^3{k}}{(2\pi)^3} {{v}_{i}^{2}}\left(-\frac{\partial{f_{\textnormal{eq}}}}{\partial{{\epsilon}_{\mathbf{k}}}}\right),
\end{equation}
where we know that $(-{\partial{f_{\textnormal{eq}}}}/{\partial{{\epsilon}_{\mathbf{k}}}})=\delta(\mu-{\epsilon}_{\mathbf{k}})$ in the zero temperature limit.  The semiclassical band velocities of the system can be obtained as %
\begin{equation}
v_{x}=\frac{2\alpha_{2}^{2}k^{2}\sin^3{\theta}\cos\phi}{\hbar\sqrt{{\alpha_{2}^{2}}k^{2}\sin^4{\theta}+\nu_{z}^{2}\cos^2{\theta}}},
\end{equation}
\begin{equation}
v_{y}=\frac{2\alpha_{2}^{2}k^{2}\sin^3{\theta}\sin\phi}{\hbar\sqrt{{\alpha_{2}^{2}}k^{2}\sin^4{\theta}+\nu_{z}^{2}\cos^2{\theta}}},
\end{equation}
and 
\begin{equation}
v_{z}=\frac{\nu_{z}^{2}\cos{\theta}}{\hbar\sqrt{{\alpha_{2}^{2}}k^{2}\sin^4{\theta}+\nu_{z}^{2}\cos^2{\theta}}}.
\end{equation}
The Drude conductivity along the $x$ direction can be obtained as:
\begin{equation} \label{drudex}
{\sigma}_{xx}^{(0)}=\frac{e^2\tau}{(2\pi)^3}\int_{0}^{\pi}\sin\theta d\theta\int_{0}^{2\pi}d\phi\int_{0}^{\infty}{k^2}dk {v_{x}^{2}}{\delta(\mu-{\epsilon}_{\mathbf{k}})}.
\end{equation}
We have used the property of Dirac delta function to evaluate the above integrand which is given by 
\begin{equation}
\delta[g(x)]=\sum_{i}\frac{\delta(x-{x_i})}{{|g^{'}({x_i})|}},
\end{equation}
where ${x_i}$ are the roots or zeroes of function $g(x)$. For the case of quadratic TCFs, using the above equation, $\delta(\mu-\epsilon_{\mathbf{k}})\equiv \delta[{g(k)}] $ can be written as
\begin{equation}
\delta\left[\mu-\sqrt{{\alpha_{2}^{2}}k^{4}\sin^4{\theta}+\nu_{z}^{2}{k^2}\cos^2{\theta}}\right]=\frac{\delta(k-{k_F                                                                                         })}{{|g^{'}({k_F})|}},
\end{equation} 
where the root of $g(k)$ for the conduction band is obtained as
\begin{equation}
k_F={\left(\frac{-{\nu_{z}^{2}}\cos^{2}\theta+{\sqrt{\nu_{z}^{4}\cos^4{\theta}+4{\alpha_{2}^{2}}\mu^{2}\sin^4{\theta}}}}{2\alpha_{2}^{2}\sin^4{\theta}}\right)}^{1/2}.
\end{equation}
Using the expresion of $k_F$ in eq. (\ref{drudex}), we get
\begin{widetext}

\begin{equation}{\label{drude-xx}}
{\sigma}_{xx}^{(0)}=2\left(\frac{e^2\tau{\nu_{z}^{3}}}{8\sqrt{2}{\pi^2}{\alpha_{2}^{2}}{\hbar^{2}}}\right)\int_{0}^{\pi} d\theta
\left(\frac{\left({-\cos^2{\theta}+\sqrt{\cos^4{\theta}+4\tilde{\mu}^{2}\sin^4{\theta}}}\right)^3}{\sin^{5}{\theta}\l(\sqrt{\cos^4{\theta}+4\tilde{\mu}^{2}\sin^4{\theta}}\r) \l(\sqrt{\cos^2{\theta}+\sqrt{\cos^4{\theta}+4\tilde{\mu}^{2}\sin^4{\theta}}}\r)}\right) 
\end{equation}
and ${\sigma}_{yy}^{(0)}={\sigma}_{xx}^{(0)}$. The Drude conductivity along the $z$ direction is obtained as
\begin{equation}{\label{drude-zz}}
{\sigma}_{zz}^{(0)}=2\left(\frac{e^2\tau{\nu_{z}^{3}}}{8\sqrt{2}{\pi^2}{\alpha_{2}^{2}}{\hbar^{2}}}\right)\int_{0}^{\pi} d\theta
\left(\frac{{-\cos^2{\theta}+\sqrt{\cos^4{\theta}+4\tilde{\mu}^{2}\sin^4{\theta}}}}{\sin^{3}{\theta}\sqrt{\cos^4{\theta}+4\tilde{\mu}^{2}\sin^4{\theta}}}\right)\left(\frac{2\cos^2{\theta}}{{\sqrt{\cos^2{\theta}+\sqrt{\cos^4{\theta}+4\tilde{\mu}^{2}\sin^4{\theta}}}}}\right).
\end{equation}
\end{widetext}
In the above expressions, an overall factor of 2 is  multiplied, covering the contributions from two triple-component nodes. We have plotted the numerically calculated Drude conductivities as a function of the Fermi energy in Fig. \ref{003}(a) and \ref{003}(b). The large difference in the values of ${\sigma}_{xx}^{(0)}$ and ${\sigma}_{zz}^{(0)}$ illustrates the strong anisotropic nature of the system.  
The variation of ${\sigma}_{zz}^{(0)}$ is more linear in Fermi energy than ${\sigma}_{xx}^{(0)}$.
Next, we consider the two orientations of magnetic field i.e., ${\mathbf{B}}$ in the $x$-$y$ plane and ${\mathbf{B}}$ in the $x$-$z$ plane. We have computed numerically the different components of electrical conductivity for the low-energy model of the quadratic TCFs (including the contribution of two triple-component nodes) in the following subsections.

\begin{figure}[htbp]
\includegraphics[trim={0cm 0cm 0cm  0cm},clip,width=8.5
cm]{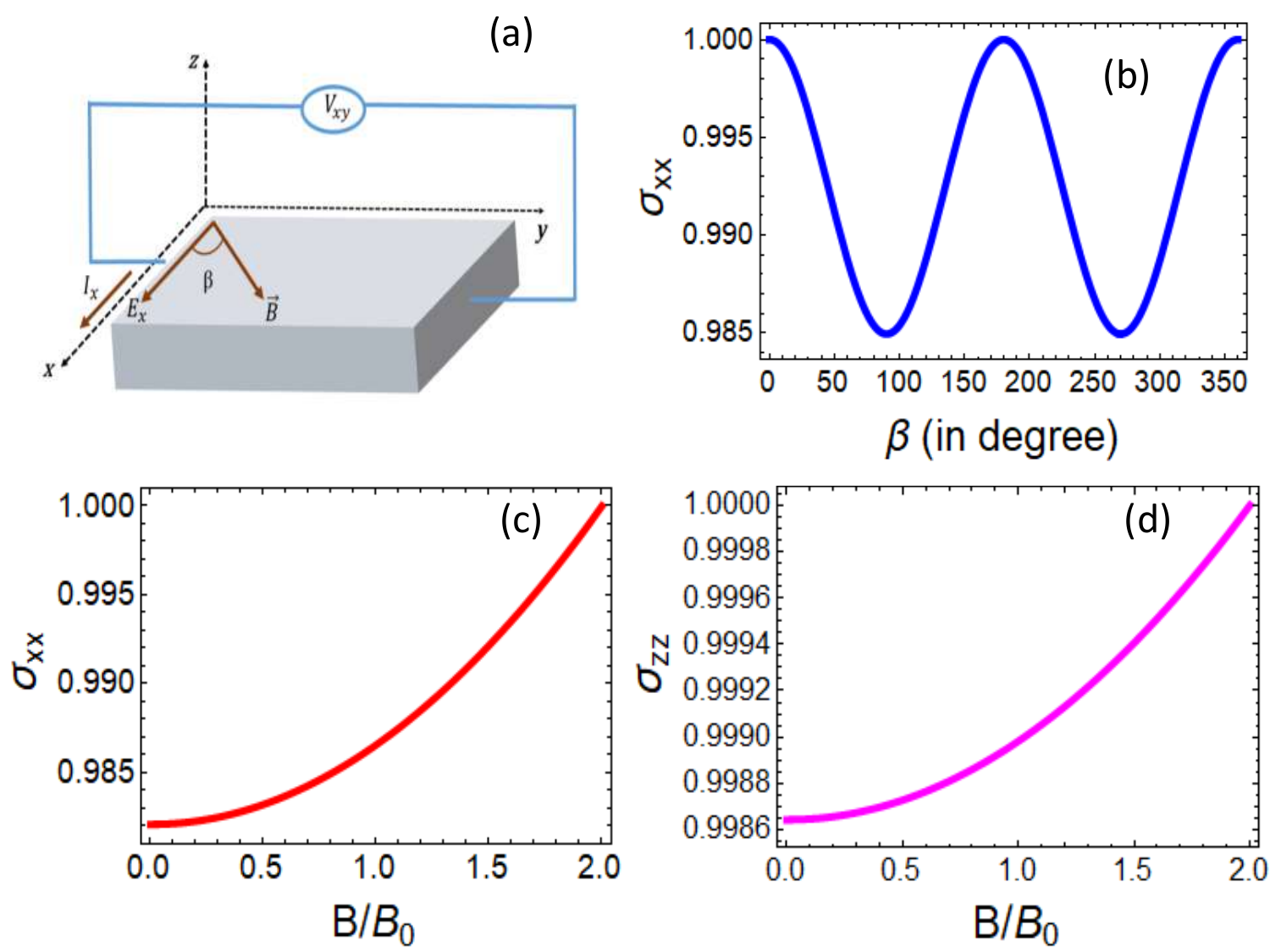}
\caption{(a) Schematic illustration of the planar Hall effect geometry. The electric field ($\mathbf{E}$) is applied along the $x$-axis and the magnetic field ($\mathbf{B}$) is applied in the $x$-$y$ plane at an angle $\beta$ from the $x$-axis. The $V_{xy}$  is measured as an in-plane induced voltage perpendicular to the direction of the  electric field. (b) The angular dependence of longitudinal magnetoconductivity ($\sigma_{xx} \propto\cos^2\beta)$ computed numerically from the  low-energy model of the quadratic TCFs at  $B/B_{0}=2$, where the $y$ axis is normalized by $\sigma_{xx}(\beta=0)$. (c) and (d) depicts the variation of LMC at $\beta=0$ (normalized by $\sigma_{xx}$ for $B/B_{0}=2$) and the out-of-plane magnetoconductivity  $\sigma_{zz}$ (normalized by $\sigma_{zz}$ for $B/B_{0}=2$) as a function of magnetic field and their amplitude in both the cases shows $B^2$ dependence respectively. Here, we have used the parameter: $\mu/\mu_{0}=10$.}
\label{01}
\end{figure}

\begin{center}
\textbf{A. Magnetic field applied in the $x$-$y$ plane}
\end{center}
We consider a planar Hall geometry where the applied electric field is along the $x$ axis (${\mathbf{E}}=E{\hat{x}}$) and the magnetic field is  confined in the $x$-$y$ plane making an angle $\beta$ with the $x$ axis, i.e., ${\mathbf{B}}= (B\cos\beta, B\sin\beta, 0)$ as shown in Fig. \ref{01}(a). 
The electric and magnetic fields are parallel to each other at $\beta=0$ resulting in maximum value of ${\mathbf{E}}\cdot{\mathbf{B}}$.

The expression for the longitudinal magnetoconductivity using Eq. (\ref{sigma_ij}) can be written as
\begin{equation}{\label{sigma_xx}}
\sigma_{xx}  = e^{2}\tau\int\frac{d^3{k}}{(2\pi)^3}{D}_{\mathbf{k}}^{-1}\bigg[{{v}}_{x}+\frac{e{B\cos\beta}}{\hbar}
({{\mathbf{v}}_{\mathbf{k}}}\cdot \bm{\Omega_\mathbf{k}})\bigg]^{2}
 \bigg(-\frac{{\partial{{f}_{\textnormal{eq}}}}}{\partial{{\epsilon}_{\mathbf{k}}}} \bigg),
\end{equation}
Further, the magnetoconductivity in the planar configuration $\sigma_{xx}$ can be  expressed explicitly in terms of  ${\sigma}_{\parallel}$ and ${\sigma}_{\perp}$ as
\begin{equation}{\label{LMC}}
{\sigma}_{xx} ={\sigma}_{\perp}+\Delta{\sigma}\cos^{2}\beta,
\end{equation}
where $\Delta{\sigma}={\sigma}_{\parallel}-{\sigma}_{\perp}$ with ${\sigma}_{\parallel}= {\sigma}_{xx}(\beta=0)={\sigma}_{xx}^{(0)}+{\sigma}_{\parallel}^{(2)}$ and ${\sigma}_{\perp}={\sigma}_{xx}(\beta={\pi}/2)= {\sigma}_{xx}^{(0)}+{\sigma}_{\perp}^{(2)}$. The $\cos^{2}\beta$ results in oscillations in the longitudinal magnetoconductivity  when the magnetic field is rotated in the $x-y$ plane, as  shown in Fig. \ref{01}(b). The $B^2$ dependence of LMC is depicted in  Fig. \ref{01}(c) for $\beta=0$.

\begin{figure}[htbp]
\includegraphics[trim={0cm 7.5cm 0cm  1cm},clip,width=8.5
cm]{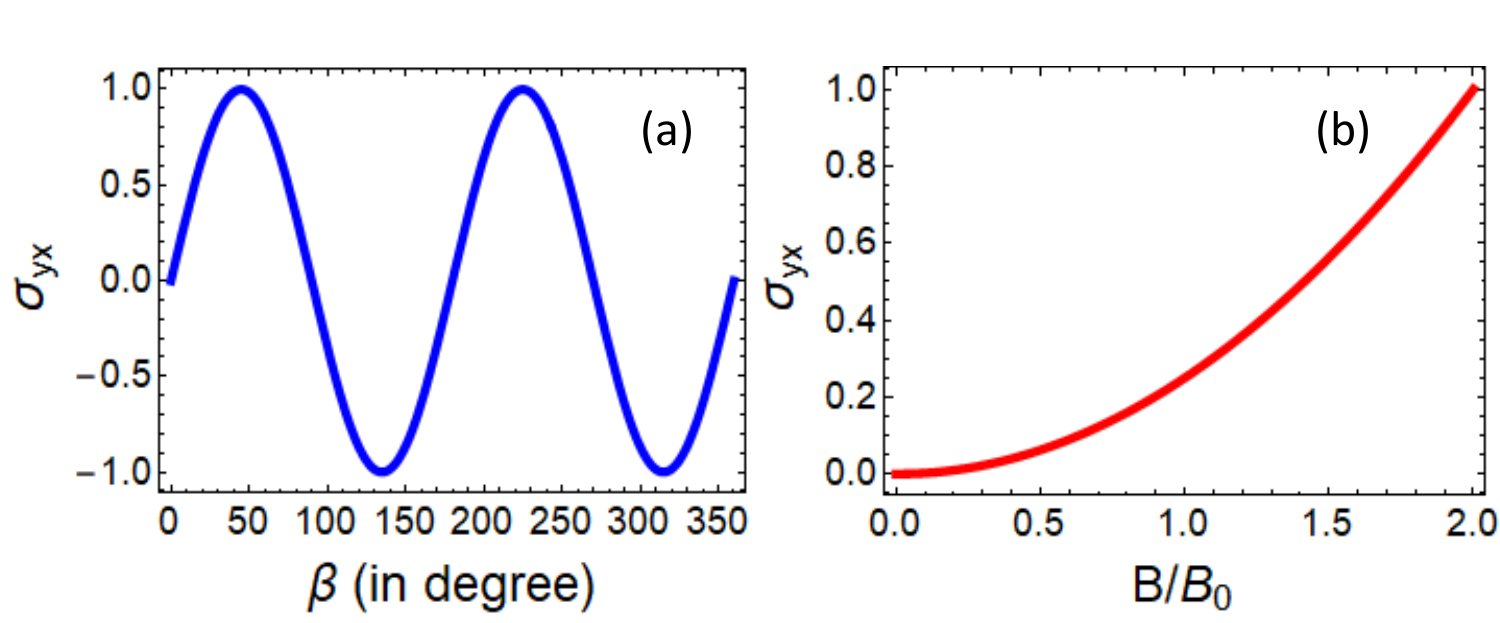}
\caption{(a) Plot of the angular behavior of planar Hall conductivity $\sigma_{yx}$ (normalized by $\sigma_{yx}(\beta=\pi/4)$) at $B/B_{0}=2$  for the quadratic TCFs with two triple-component nodes. PHC follows the standard angular trend of $\sin\beta\cos\beta$. (b) The amplitude of PHC (normalised by its amplitude for $B/B_{0}=2$) as a function of magnetic field. PHC shows $B^2$ dependence on the magnetic field. We have taken $\mu/\mu_{0}=10$. }
\label{02}
\end{figure}

\textbf{\textit{Planar Hall effect}}: We observe planar Hall effect (PHE) in the quadratic TCFs for the given planar geometry. 
The PHE\cite{PHE1,PHE2} refers to a phenomenon where a voltage is induced perpendicular to the applied electric field in presence of a magnetic field which is coplanar with the applied field and the induced voltage.  The expression for planar Hall conductivity (PHC) is obtained as
\begin{equation}{\label{sigma_yx}}
\begin{centering}
\begin{aligned}
\sigma_{yx} & = e^{2}\tau\int\frac{d^3{k}}{(2\pi)^3}{({D}_{\mathbf{k}})}^{-1}\bigg[{{v}}_{x}+\frac{e{B\cos\beta}}{\hbar}
({{\mathbf{v}}_{\mathbf{k}}}\cdot \bm{\Omega_\mathbf{k}})\bigg]\\
& \bigg[{{v}}_{y}+\frac{e{B\sin\beta}}{\hbar}
({{\mathbf{v}}_{\mathbf{k}}}\cdot \bm{\Omega_\mathbf{k}})\bigg]\bigg(-\frac{{\partial{{f}_{\textnormal{eq}}}}}{\partial{{\epsilon}_{\mathbf{k}}}} \bigg),
\end{aligned}
\end{centering}
\end{equation}
Using the above expression, the PHC can be written as
\begin{equation}
{\sigma}_{yx}= \Delta{\sigma}\sin\beta \cos\beta,
\end{equation}
where $\Delta\sigma$ has been defined after Eq. (\ref{LMC}). 
The angular dependence of PHC is shown in Fig. \ref{02}(a). The PHC reaches the maximum value at an odd multiple of $\pi/4$. The amplitude of PHC ($\Delta\sigma$) shows a quadratic dependence on $B$ for any value of $ \beta $ (except for $\beta = 0$ and $\beta = \pi/2$ where it trivially vanishes) as shown in Fig. \ref{02}(b).  Since its origin is associated with Berry curvature and not the Lorentz force, the PHC obeys a symmetric relation ${\sigma}_{yx}={\sigma}_{xy}$ unlike the regular Hall conductivity.
The other diagonal elements of conductivity for different directions of electric field are calculated as: ${\sigma}_{yy}(\beta)={\sigma}_{xx}(\pi/2-\beta)$ and the zz-component of conductivity is given by
\begin{equation}{\label{sigma_zz}}
\sigma_{zz}  = e^{2}\tau\int\frac{d^3{k}}{(2\pi)^3}({D}_{\mathbf{k}}^{-1}){{v}}_{z}^{2}
 \bigg(-\frac{{\partial{{f}_{\textnormal{eq}}}}}{\partial{{\epsilon}_{\mathbf{k}}}} \bigg),
\end{equation}

It also follows quadratic-$B$ dependence as shown in Fig. \ref{01}(d) and no angular dependence. It is to be noted that the lowest-order  magnetic field correction to the conductivity comes quadratic and  the $B$-linear dependence of conductivity is zero in our system. All the other off-diagonal components of conductivity are calculated to be zero.

 \begin{figure}[htbp]
\includegraphics[trim={0cm 0cm 0cm  9cm},clip,width=8.5
cm]{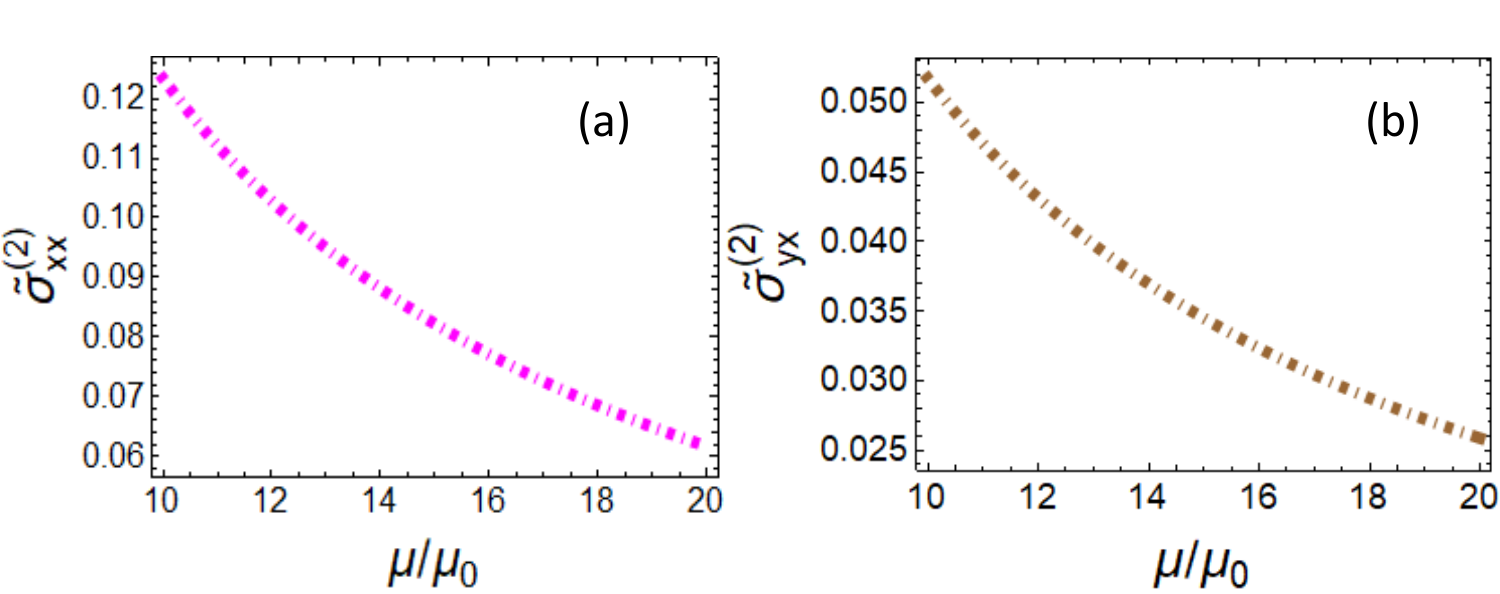}
\caption{(a)-(b) depicts the behavior of quadratic $B$-correction to longitudinal magnetoconductivity (at $\beta=0$) and planar Hall conductivity (at $\beta=\pi/4$) with the Fermi energy at $B/B_{0}=2$. The quadratic in $B$ component of LMC and PHC shows decrease with the Fermi energy.} 
\label{03}
\end{figure}

\textbf{\textit{Dependence on the Fermi energy} ($\mu/\mu_{0}$)}:  In Fig. \ref{03}(a) and \ref{03}(b), the variation of quadratic in $B$ component of the longitudinal magnetoconductivity ($\sigma_{xx}^{(2)}$) and planar Hall conductivity ($\sigma_{yx}^{(2)}$) is plotted with respect to the Fermi energy. Both of them decrease with the Fermi energy. This is expected because the magnitude of Berry curvature  decreases as the Fermi energy shifts away from the band touching node.
\begin{center}
\textbf{B. Magnetic field applied in the $x$-$z$ plane}
\end{center}
Now, we consider another planar Hall setup where the electric field is applied along the $z$ direction (${\mathbf{E}}=E{\hat{z}}$) and the magnetic field is rotated in the $x$-$z$ plane such that it makes an angle $\beta$ with respect to the $z$ axis, i.e., ${\mathbf{B}}= (B\sin\beta, 0, B\cos\beta)$. 

\begin{figure}[htbp]
\includegraphics[trim={0cm 0cm 0cm  0cm},clip,width=8.5
cm]{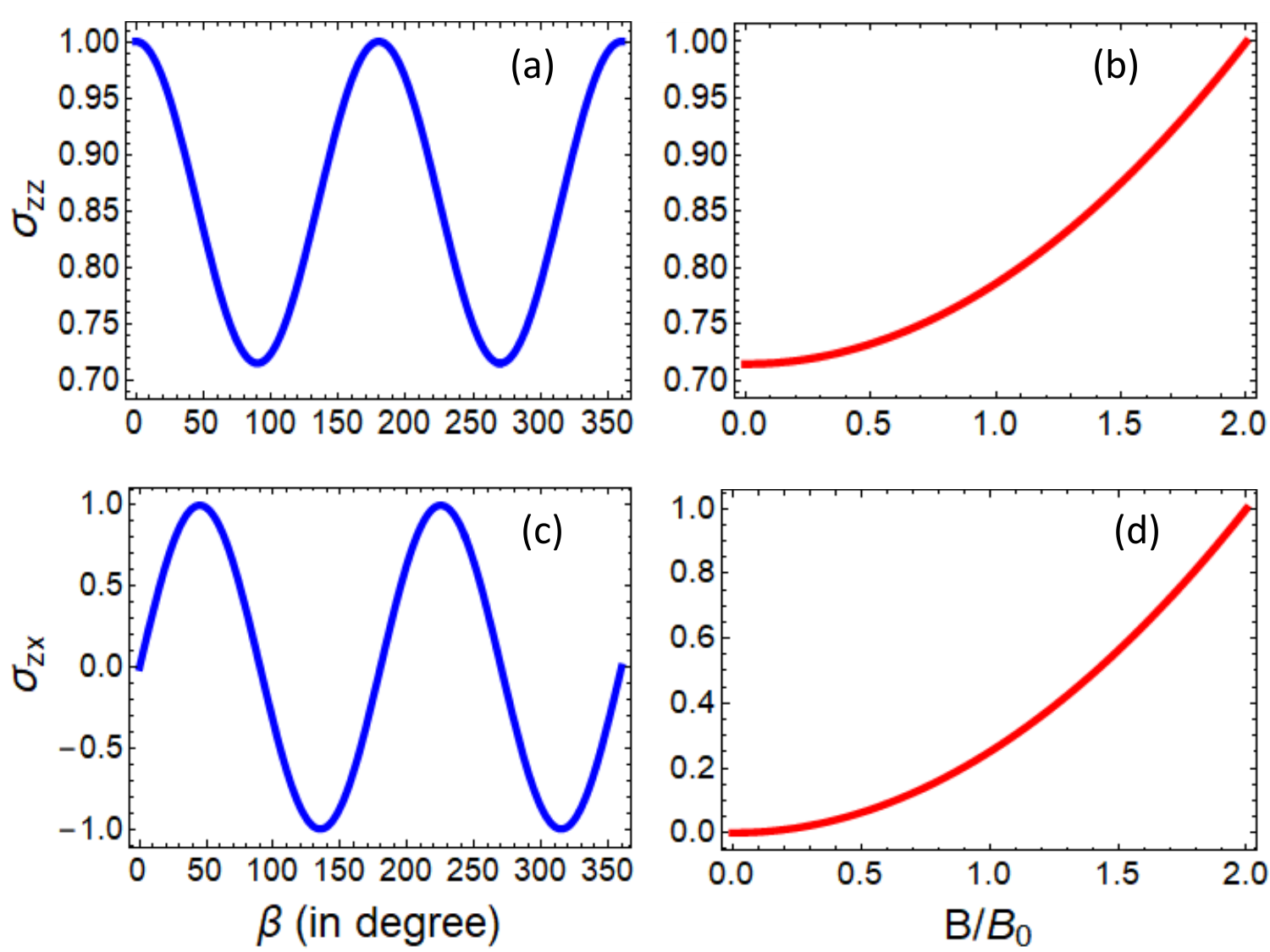}
\caption{(a) and (c) shows the angular dependence of longitudinal magnetoconductivity ($\sigma_{zz} \propto\cos^2\beta)$ and planar Hall conductivity ($\sigma_{zx} \propto\sin\beta\cos\beta$) calculated numerically for the quadratic TCFs, when a magnetic field is rotated in the $x$-$z$ plane at $B/B_{0}=2$. Here, the $y$ axes of (a) and (c) are normalized by $\sigma_{zz}(\beta=0)$ and $\sigma_{zx}(\beta=\pi/4)$ respectively. (b) and (d) depicts the variation of LMC at $\beta=0$ (normalized by $\sigma_{zz}$ for $B/B_{0}=2$) and the amplitude of planar Hall conductivity (normalized by $\sigma_{zx}$ for $B/B_{0}=2$) as a function of magnetic field. The calculations are performed at $\mu/\mu_{0}=10$.} 
\label{04}
\end{figure}

\begin{figure}[htbp]
\includegraphics[trim={0cm 6cm 0cm  0cm},clip,width=8.5
cm]{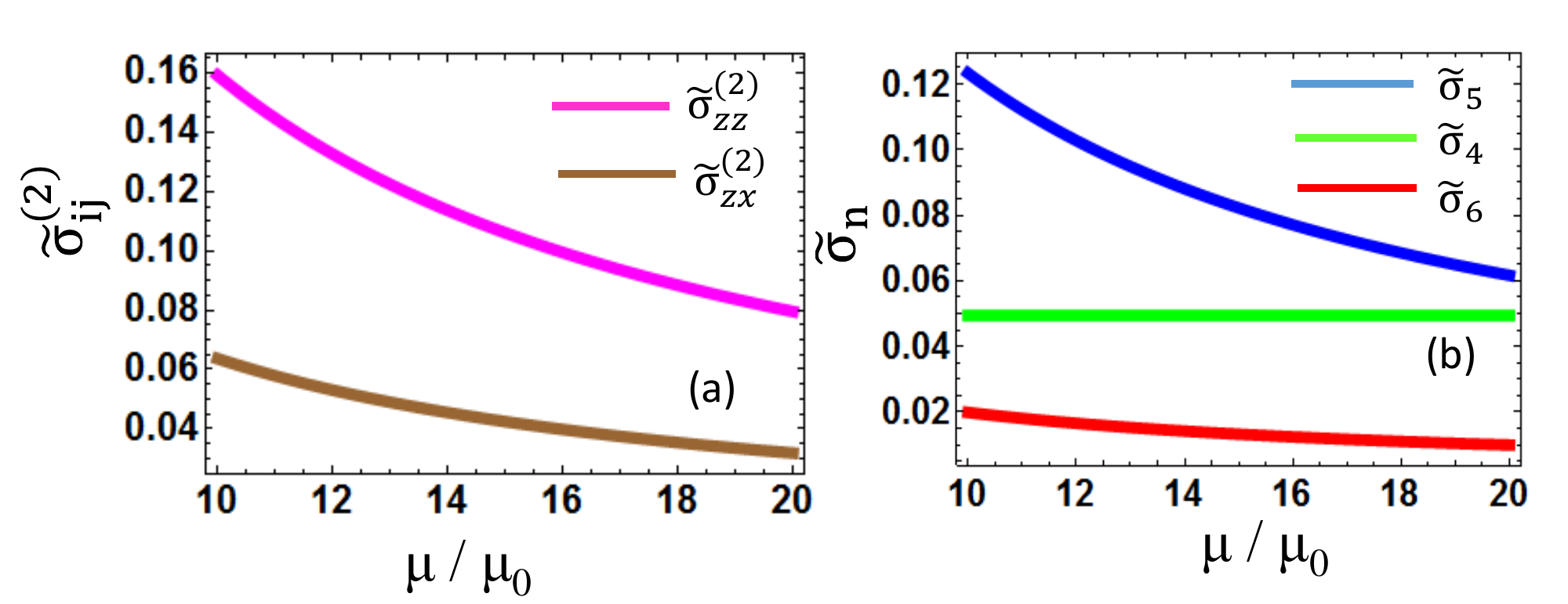}
\caption{(a) Plots of variation of quadratic $B$-correction to longitudinal magnetoconductivity (${\sigma}_{zz}^{(2)}$ at $\beta=0$) and planar Hall conductivity (${\sigma}_{zx}^{(2)}$ at $\beta=\pi/4$) with the Fermi energy for $B/B_{0}=2$ respectively. In both the cases, the quadratic in $B$ component of LMC and PHC shows decrease with the Fermi energy. (b) Dependence of the quadratic-$B$ correction in  ${\sigma}_{xx}$ and ${\sigma}_{yy}$ when ${\bf B}\ ||\ \hat{\bf x}$, i.e., (${\sigma}_{5}$ and ${\sigma}_{6}$) with the Fermi energy. Both the conductivity coefficients decreases with the Fermi energy. The quadratic-$B$ correction in both $\sigma_{xx}$ and $\sigma_{yy}$ when ${\bf B}\ ||\ \hat{\bf z}$ denoted by ${\sigma}_{4}$ is independent of the Fermi energy. In the plot, ${\sigma}_{4}$ is divided by a factor of 10. }  
\label{05}
\end{figure}

The longitudinal magnetoconductivity for the above configuration using Eq. (\ref{sigma_ij}) can be written as
\begin{equation}{\label{sigma_zz1}}
\sigma_{zz}  = e^{2}\tau\int\frac{d^3{k}}{(2\pi)^3}{D}_{\mathbf{k}}^{-1}\bigg[{{v}}_{z}+\frac{e{B\cos\beta}}{\hbar}
({{\mathbf{v}}_{\mathbf{k}}}\cdot \bm{\Omega_\mathbf{k}})\bigg]^{2}
 \bigg(-\frac{{\partial{{f}_{\textnormal{eq}}}}}{\partial{{\epsilon}_{\mathbf{k}}}} \bigg).
\end{equation} 
The above expression takes the following form
\begin{equation}{\label{lmc}}
{\sigma}_{zz} ={\sigma}_{zz}^{(0)}+{\sigma}_{1}+{\sigma}_{2}\cos^{2}\beta,
\end{equation}
where ${\sigma}_{zz}^{(0)}$ is the Drude conductivity  given by Eq. (\ref{drude-zz}) and the conductivity coefficients ${\sigma}_{1}$ and ${\sigma}_{2}$ are proportional to $B^2$. All the three terms of ${\sigma}_{zz}$ show significant dependence on Fermi energy.  The longitudinal magnetoconductivity shows the angular dependence of $\cos^{2}\beta$ as shown in Fig. \ref{04}(a). Its $B^2$ dependence (except at $\beta=\pi/2$) is depicted in Fig. \ref{04}(b).
The simplified expression of the planar Hall conductivity for $B$ applied in the $x$-$z$ plane can be written as
\begin{equation}{\label{sigma_zx}}
\begin{centering}
\begin{aligned}
\sigma_{zx} & = e^{2}\tau\int\frac{d^3{k}}{(2\pi)^3}{({D}_{\mathbf{k}})}^{-1}\bigg[{{v}}_{z}+\frac{e{B\cos\beta}}{\hbar}
({{\mathbf{v}}_{\mathbf{k}}}\cdot \bm{\Omega_\mathbf{k}})\bigg]\\
& \bigg[{{v}}_{x}+\frac{e{B\sin\beta}}{\hbar}
({{\mathbf{v}}_{\mathbf{k}}}\cdot \bm{\Omega_\mathbf{k}})\bigg]\bigg(-\frac{{\partial{{f}_{\textnormal{eq}}}}}{\partial{{\epsilon}_{\mathbf{k}}}} \bigg),
\end{aligned}
\end{centering}
\end{equation}
We can also write the PHC as
\begin{equation}
{\sigma}_{zx}= {\sigma}_{3}\sin\beta \cos\beta.
\end{equation}
Its angular and quadratic-$B$ dependences are shown in Fig. \ref{04}(c) and  Fig. \ref{04}(d) respectively. The maximum value is obtained at $\beta=\pi/4$.  
The variations of quadratic in $B$ component of the longitudinal magnetoconductivity ($\sigma_{zz}^{(2)}$) and  planar Hall conductivity ($\sigma_{zx}$) with respect to the Fermi energy are plotted in Fig. \ref{05}(a). Similar to the previous orientation, both the quantities decrease with Fermi energy.

The other diagonal components of the conductivity are obtained as 
\begin{equation}{\label{xx}}
{\sigma}_{xx} ={\sigma}_{xx}^{(0)}+{\sigma}_{4}\cos^{2}\beta+{\sigma}_{5}\sin^{2}\beta
\end{equation}
and
\begin{equation}{\label{yy}}
{\sigma}_{yy} ={\sigma}_{yy}^{(0)}+{\sigma}_{4}\cos^{2}\beta+{\sigma}_{6}\sin^{2}\beta.
\end{equation}

Here, ${\sigma}_{xx}^{(0)}={\sigma}_{yy}^{(0)}$, calculated numerically from Eq. (\ref{drude-xx}). Apart from the Drude parts, there are quadratic-$B$ corrections in conductivity viz. ${\sigma}_{4}$, ${\sigma}_{5}$ and ${\sigma}_{6}$. Here, $\sigma_4$ denotes the quadratic-$B$ correction in both $\sigma_{xx}$ and $\sigma_{yy}$ when ${\bf B}\ ||\ \hat{\bf z}$ while $\sigma_5$ and $\sigma_6$ denotes the quadratic-$B$ correction in $\sigma_{xx}$ and $\sigma_{yy}$ respectively when ${\bf B}\ ||\ \hat{\bf x}$. The variation of these correction terms with Fermi energy is shown in Fig. \ref{05}(b). We observe that $\sigma_4$ is independent of Fermi energy while  $\sigma_5$ and $\sigma_6$ show a decrease. 

An important feature which distinguishes this orientation of planar Hall geometry from the previous one is that the out-of-plane component of conductivity $\sigma_{yy}$ also has an oscillatory behaviour as the magnetic field is rotated in the $x$-$z$ plane.  
\section{Magnetoresistance}\label{V}
We have evaluated the magnetoconductivity matrix numerically from the low-energy
model of the quadratic TCFs  for two different planar geometries in the last section.  Next, we calculate the magnetoresistance  which is defined as 
\begin{equation}{\label{MR}}
\textnormal{MR}_{ii} = \frac{{\rho_{ii}(B)}-{\rho_{ii}(0)}}{{\rho_{ii}(0)}}, 
\end{equation}
where $\rho_{ii}$ denotes the diagonal components of the resistivity tensor with $i = x, y, z$. The Drude resistivity $\rho_{ii}(0)$ is simply the inverse of Drude conductivity $\sigma_{ii}^{(0)}$.  
In this section also, we study and compare magnetoresistance for the two orientations of magnetic field mentioned earlier.

\begin{figure}[htbp]
\includegraphics[trim={0cm 0cm 0cm  0cm},clip,width=8.5
cm]{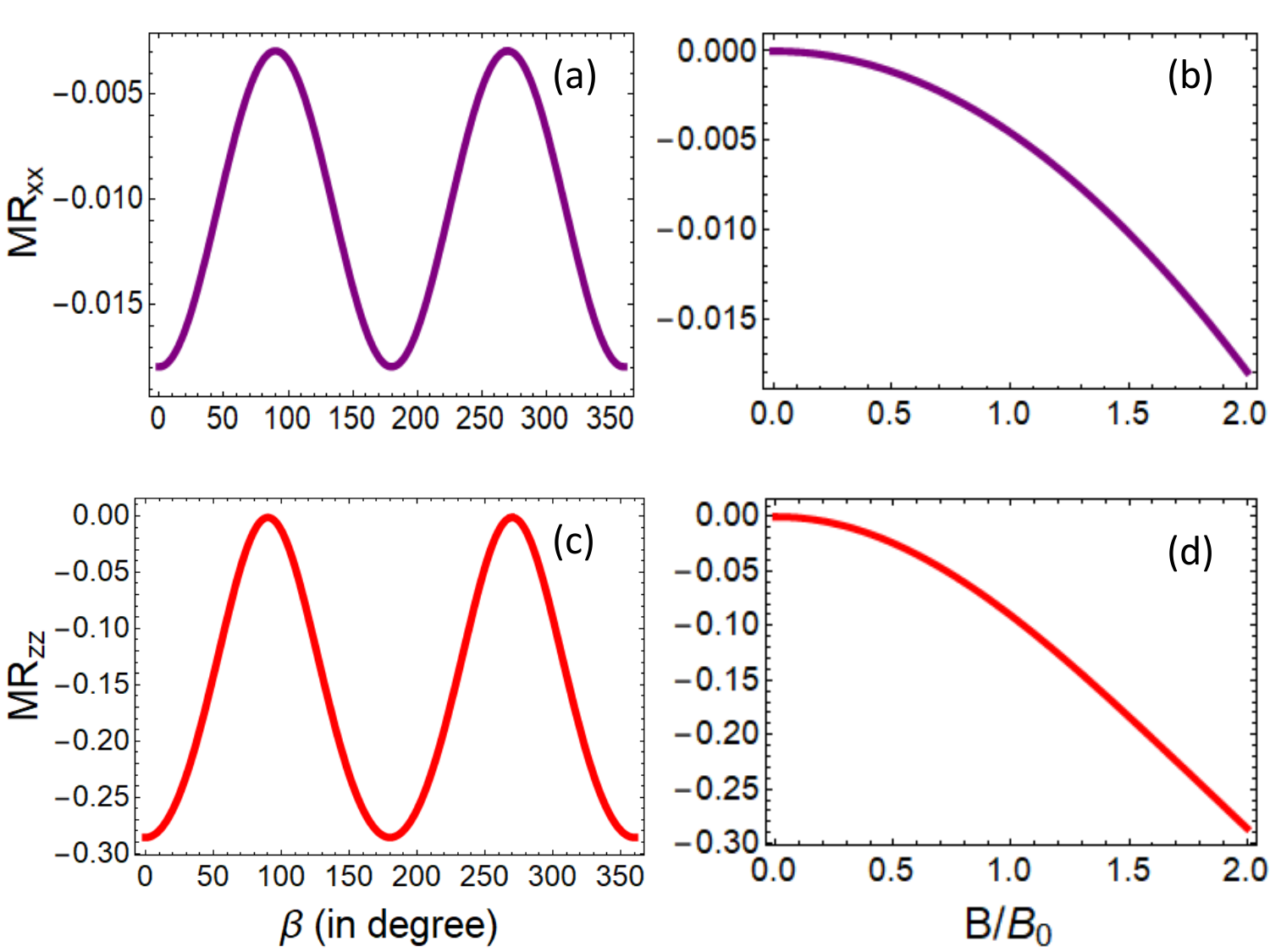}
\caption{The angular and magnetic field dependence of magnetoresistance for quadratic TCFs evaluated for two different configurations of the magnetic field. Here, the purple curve represents the case of magnetic field applied in the $x$-$y$ plane whereas red curve denotes the magnetic field confined in the $x$-$z$ plane. The planar MR [$\textnormal{MR}_{xx}(\beta)$] and [$\textnormal{MR}_{zz}(\beta)$] varies as $\cos^{2}\beta$ corresponding to their respective magnetic field configuration at $B/B_{0}=2$ is shown in (a) and (c). The variation of longitudinal magnetoresistance: (b) $\textnormal{MR}_{xx}$ and (d) $\textnormal{MR}_{zz}$  as a function of magnetic field at $\beta=0$. We have used $\mu/\mu_{0}=10$. } 
\label{06}
\end{figure}

For the case of magnetic field applied in the $x$-$y$ plane making an angle $\beta$ with the $x$ axis, the planar MR [$\textnormal{MR}_{xx}(\beta)$] is anisotropic as it shows an angular variation of $\cos^{2}\beta$ shown in Fig. \ref{06}(a). The decrease in magnetoresistivity is maximum at $\beta= 0$ and $\pi$ and minimum at $\beta= \pi/2$. It is evident from Fig. \ref{06}(a), when $\mu/\mu_{0}=10$ and $B/B_{0}=2$, the negative MR resulting from the Berry curvature is about $-1.8\%$ at $\beta= 0$ and $\pi$. We have plotted the magnetic field dependence of longitudinal MR ($\textnormal{MR}_{xx}$) at $\beta=0$  in Fig. \ref{06}(b).
For the second orientation where applied magnetic field is rotated in the $x$-$z$ plane by making an angle  $\beta$ with the $z$ axis, the planar MR is denoted by [$\textnormal{MR}_{zz}(\beta)$] which also follows the angular dependence of $\cos^{2}\beta$  as shown in Fig. \ref{06}(c). Here, the MR reaches about $-29\%$ and thus the Berry curvature effects on MR are considerably large in this orientation. The increasing nature of absolute longitudinal MR ($\textnormal{MR}_{zz}$) at $\beta=0$  with the magnetic field is shown in Fig. \ref{06}(d).

\begin{figure}[htbp]
\includegraphics[trim={0cm 5cm 0cm  0cm},clip,width=8.5
cm]{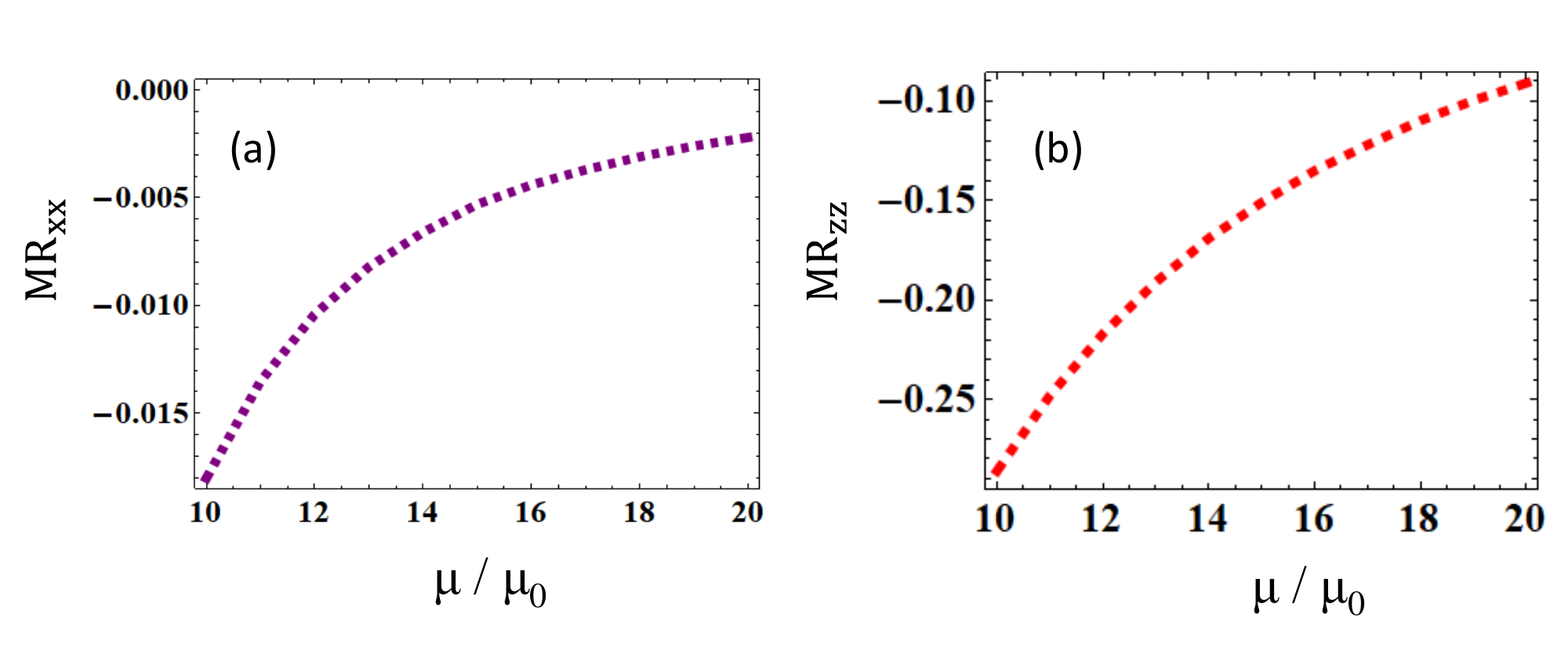}
\caption{The variation of magnetoresistance with the Fermi energy. (a) The longitudinal MR (MR$_{xx}$) corresponds to the case of magnetic field applied in the $x$-$y$ plane (purple dashed curve) and (b)  MR$_{zz}$ is the longitudinal MR for the magnetic field confined in the $x$-$z$ plane (red dashed curve). In both the cases, the absolute value of MR decreases with the Fermi energy. We have used the following parameters: $B/B_{0}=2$ and $\beta=0$. } 
\label{07}
\end{figure}

\textbf{\textit{Dependence on the Fermi energy} ($\mu/\mu_{0}$)}: The longitudinal magnetoresistances MR$_{xx}$ and MR$_{zz}$ corresponding to their respective magnetic field configurations are plotted as function of Fermi energy in Fig. \ref{07}(a) and \ref{07}(b). The absolute value of MR in both the cases decrease with the Fermi energy.

\section{Thermal conductivity}\label{VI}
 
The thermal current is defined as\cite{ashcroft,PTHE1}
\begin{equation}{\label{j_thermal}}
{\mathbf{j}}_{q} = \int[d\mathbf{k}]{D}_{\mathbf{k}}({\epsilon}_{\mathbf{k}}-\mu){\dot{\mathbf{r}}}{f}_{{\mathbf{k}}},
\end{equation}
Substituting  ${\dot{\mathbf{r}}}$ and ${f}_{\mathbf{k}}$ from  Eqs. (\ref{EOM1}) and (\ref{NDF}) in Eq. (\ref{j_thermal}) yields the following general expression of Berry curvature dependent thermal conductivity: 
\begin{equation}{\label{kappa_ij}}
\begin{centering}
\begin{aligned}
\kappa_{ij} & = -\frac{1}{\hbar}\int[d\mathbf{k}]\epsilon_{ijl}\Omega_{\mathbf{k}}^{l}\frac{({\epsilon}_{\mathbf{k}}-\mu)^{2}}{T}{f}_{\textnormal{eq}}\\
&+\tau\int[d\mathbf{k}]{({D}_{\mathbf{k}})}^{-1}\bigg({{v}}_{i}+\frac{e}{\hbar}{B_{i}}
({{\mathbf{v}}_{\mathbf{k}}}\cdot \bm{\Omega_\mathbf{k}})\bigg)\\
& \bigg({{v}}_{j}+\frac{e}{\hbar}{B_{j}}
({{\mathbf{v}}_{\mathbf{k}}}\cdot \bm{\Omega_\mathbf{k}})\bigg)\frac{({\epsilon}_{\mathbf{k}}-\mu)^{2}}{T}\bigg(-\frac{{\partial{{f}_{\textnormal{eq}}}}}{\partial{{\epsilon}_{\mathbf{k}}}} \bigg),
\end{aligned}
\end{centering}
\end{equation}
where the first term corresponds to the anomalous thermal Hall conductivity without magnetic field. 
The longitudinal thermal conductivity in the absence of magnetic field is obtained as
\begin{equation}{\label{kappa_ij0}}
{\kappa}_{ii}^{(0)}= \tau\int\frac{d^3{k}}{(2\pi)^3} {{v}_{i}^{2}}\frac{({\epsilon}_{\mathbf{k}}-\mu)^{2}}{T}\left(-\frac{\partial{f_{\textnormal{eq}}}}{\partial{{\epsilon}_{\mathbf{k}}}}\right).
\end{equation}
\begin{center}
\textbf{A. Magnetic field is applied in the $x$-$y$ plane}
\end{center}
We will now investigate the longitudinal magneto-thermal conductivity and planar thermal Hall conductivity for a configuration where temperature gradient is applied along the $x$ axis and magnetic field is in the $x$-$y$ plane making an angle $\beta$ from the $x$ axis such that  $\boldsymbol{\nabla}{T}={\nabla}T{\hat{x}}$, ${\mathbf{B}}= (B\cos\beta, B\sin\beta, 0)$ and ${\mathbf{E}}=0$.
 
\begin{figure}[htbp]
\includegraphics[trim={0cm 0cm 0cm  0cm},clip,width=8.5
cm]{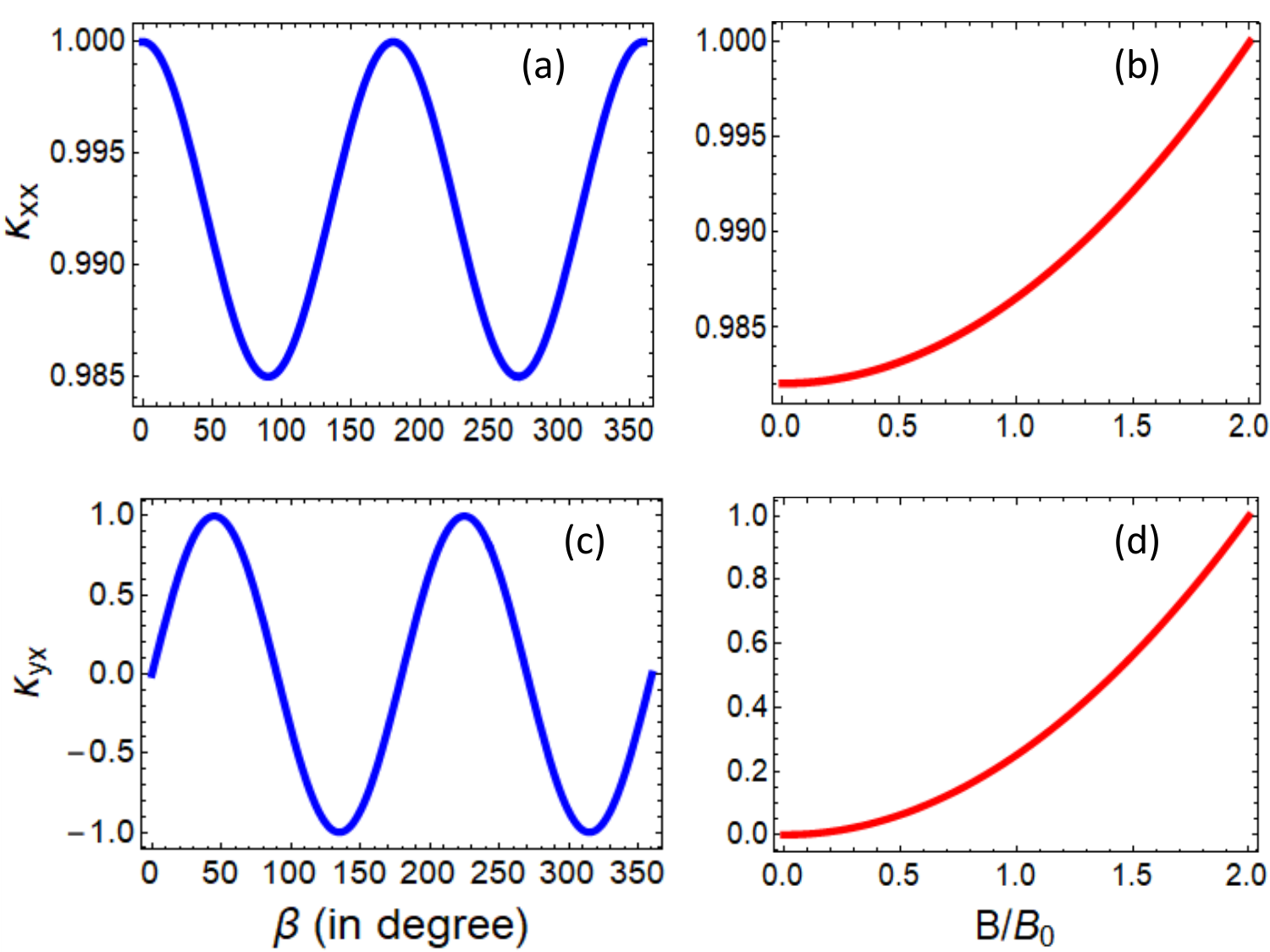}
\caption{(a) The angular dependence of longitudinal magneto-thermal conductivity $\kappa_{xx}$ (normalized by $\kappa_{xx}(\beta=0)$) computed numerically for the quadratic TCFs at $B/B_{0}=2$. (b) depicts the variation of LMTC at $\beta=0$ (normalized by $\kappa_{xx}$ for $B/B_{0}=2$) as a function of magnetic field. (c) shows the plot of the angular behavior of planar thermal Hall conductivity $\kappa_{yx}$  at $B/B_{0}=2$. (d) The amplitude of PTHC shows the quadratic-$B$ dependence.  Here, the $y$ axes of (c) and (d) are normalized by $\kappa_{yx}(\beta=\pi/4)$ and $\kappa_{yx}$($B/B_{0}=2$) respectively. The calculations are performed at $\mu/\mu_{0}=10$ and $T=300$ K.} 
\label{08}
\end{figure}
Next, we consider the temperature gradient and magnetic field are parallel in order to calculate the longitudinal magneto-thermal conductivity which is given by the following expression as
\begin{equation}{\label{kappa_xx}}
\kappa_{xx}  = \tau\int[d\mathbf{k}]\frac{({\epsilon}_{\mathbf{k}}-\mu)^{2}}{({D}_{\mathbf{k}})T}\bigg[{{v}}_{x}+\frac{e{B\cos\beta}}{\hbar}
({{\mathbf{v}}_{\mathbf{k}}}\cdot \bm{\Omega_\mathbf{k}})\bigg]^{2}
 \bigg(-\frac{{\partial{{f}_{\textnormal{eq}}}}}{\partial{{\epsilon}_{\mathbf{k}}}} \bigg).
\end{equation}
We can further express the magneto-thermal conductivity in the planar configuration $\kappa_{xx}$  in terms of  ${\kappa}_{\parallel}$ and ${\kappa}_{\perp}$ describing the cases of thermal current flowing parallel and perpendicular to the magnetic field as
\begin{equation}{\label{LMTC}}
{\kappa}_{xx} ={\kappa}_{\perp}+\Delta{\kappa}\cos^{2}\beta.
\end{equation}
Here $\Delta{\kappa}={\kappa}_{\parallel}-{\kappa}_{\perp}$ with ${\kappa}_{\parallel}= {\kappa}_{xx}(\beta=0)={\kappa}_{xx}^{(0)}+{\kappa}_{\parallel}^{(2)}$ and ${\kappa}_{\perp}={\kappa}_{xx}(\beta={\pi}/2)= {\kappa}_{xx}^{(0)}+{\kappa}_{\perp}^{(2)}$. The longitudinal magneto-thermal conductivity has the angular dependence of $\cos^2\beta$ which is depicted in Fig. \ref{08}(a) and thus shows anisotropic behavior. The LMTC shows the quadratic-$B$ dependence on the magnetic field (except at $\beta=\pi/2$) for the quadratic TCFs at $T=300$ K as shown in Fig. \ref{08}(b).

\textbf{\textit{Planar thermal Hall effect}}: The PTHC\cite{PTHE3} is defined as the appearance of in-plane transverse temperature gradient when the applied temperature gradient and magnetic field are coplanar. It is the thermal analog of planar Hall effect. The expression for planar thermal Hall conductivity  is obtained as
\begin{equation}{\label{kappa_yx}}
\begin{centering}
\begin{aligned}
\kappa_{yx} & = \tau\int\frac{d^3{k}}{(2\pi)^3}\frac{({\epsilon}_{\mathbf{k}}-\mu)^{2}}{({D}_{\mathbf{k}})T}\bigg[{{v}}_{x}+\frac{e{B\cos\beta}}{\hbar}
({{\mathbf{v}}_{\mathbf{k}}}\cdot \bm{\Omega_\mathbf{k}})\bigg]\\
& \bigg[{{v}}_{y}+\frac{e{B\sin\beta}}{\hbar}
({{\mathbf{v}}_{\mathbf{k}}}\cdot \bm{\Omega_\mathbf{k}})\bigg]\bigg(-\frac{{\partial{{f}_{\textnormal{eq}}}}}{\partial{{\epsilon}_{\mathbf{k}}}} \bigg),
\end{aligned}
\end{centering}
\end{equation}
The PHTC can be further expressed as
\begin{equation}
{\kappa}_{yx}= \Delta{\kappa}\sin\beta \cos\beta.
\end{equation}
The planar thermal Hall conductivity follows the angular dependence of $\sin\beta\cos\beta$. The amplitude of PTHC ($\Delta\kappa$) shows a $B^2$ dependence for any value of $ \beta $ except for $\beta = 0$ and $\beta = \pi/2$. We have plotted the angular and magnetic field dependence of $\kappa_{yx}$ at $T=300$ K in Fig. \ref{08}(c) and \ref{08}(d) respectively.

\begin{center}
\textbf{B. Magnetic field is applied in the $x$-$z$ plane}
\end{center}
\begin{figure}[htbp]
\includegraphics[trim={0cm 0cm 0cm  0cm},clip,width=8.5
cm]{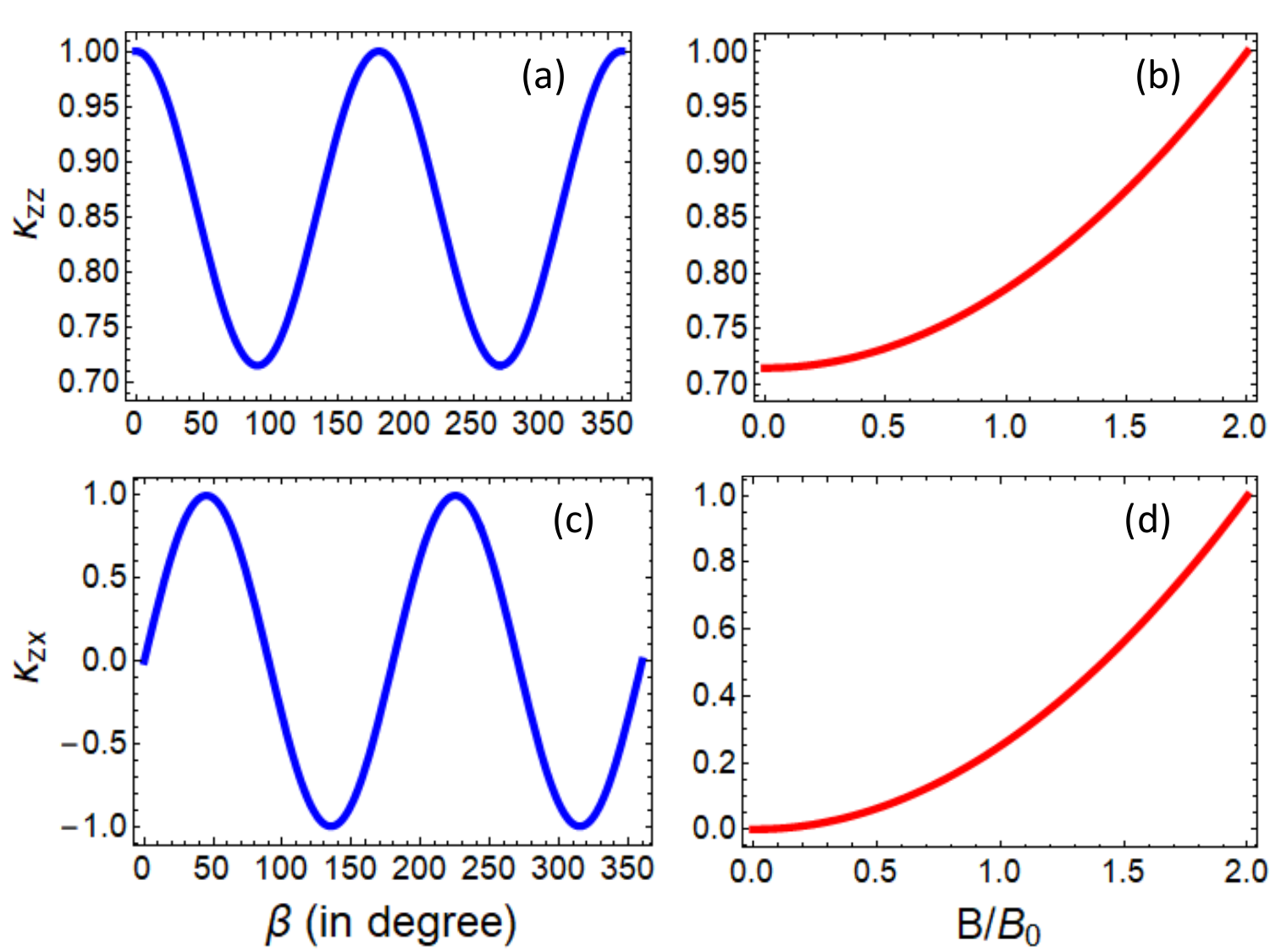}
\caption{(a) and (c) depicts the angular dependence of longitudinal magneto-thermal conductivity ($\kappa_{zz} \propto\cos^2\beta)$ and planar thermal Hall conductivity ($\kappa_{zx} \propto\sin\beta\cos\beta$) for the quadratic TCFs, when a magnetic field is rotated in the $x$-$z$ plane at $B/B_{0}=2$. Here, the $y$ axes of (a) and (c) are normalized by $\kappa_{zz}(\beta=0)$ and $\kappa_{zx}$($\beta=\pi/4)$ respectively. (b) and (d) The variation of LMTC at $\beta=0$ (normalized by $\kappa_{zz}$ for $B/B_{0}=2$) and the amplitude of planar thermal Hall conductivity (normalized by $\kappa_{zx}$ for $B/B_{0}=2$) as a function of magnetic field. We have used the parameter: $\mu/\mu_{0}=10$ and $T=300$ K.}
\label{09}
\end{figure}
Next, we fix the direction of temperature gradient along the $z$ axis and magnetic field is rotated in the $x$-$z$ plane making an angle $\beta$ from the $z$ axis and thus we have $\boldsymbol{\nabla}{T}={\nabla}T{\hat{z}}$, ${\mathbf{B}}= (B\sin\beta, 0,  B\cos\beta,)$ and ${\mathbf{E}}=0$. We have evaluated the longitudinal magneto-thermal conductivity $\kappa_{zz}$ and planar thermal Hall conductivity $\kappa_{zx}$ and discuss the numerical results of the above thermal conductivities.
The longitudinal magneto-thermal conductivity for this configuration can be expressed as
\begin{equation}{\label{lmtc}}
{\kappa}_{zz} ={\kappa}_{zz}^{(0)}+{\kappa}_{1}+{\kappa}_{2}\cos^{2}\beta.
\end{equation}
The planar thermal Hall conductivity is given by
\begin{equation}
{\kappa}_{zx}= {\kappa}_{3}\sin\beta \cos\beta.
\end{equation}
Here, the thermal conductivity coefficients (${\kappa}_{1}, {\kappa}_{2}$ and ${\kappa}_{3}$) are proportional to $B^2$. The longitudinal magneto-thermal conductivity ${\kappa}_{zz}$ has the angular dependence of $\cos^2\beta$ as depicted in Fig. \ref{09}(a), whereas the planar thermal Hall conductivity follows the standard $\sin\beta\cos\beta$ dependence which is shown in Fig. \ref{09}(c). Fig. \ref{09}(b) and \ref{09}(d) depicts the variation of ${\kappa}_{zz}$ and ${\kappa}_{zx}$ with the magnetic field for the quadratic TCFs at $T=300$ K . Both the LMTC and PTHC shows quadratic dependence on the magnetic field.

\textbf{\textit{Validity of Wiedemann-Franz law}}: The Wiedemann-Franz law states that the ratio of electronic contribution of thermal conductivity  to  electrical conductivity for  metal is proportional to the temperature and can be expressed as  $\kappa_{ij}/(\sigma_{ij}T)={L_0}$, where $L_{0}=\frac{\pi^{2}{K_{B}^{2}}}{3e^{2}}$ is the Lorentz number. The Wiedemann-Franz law is valid in our case of TRS broken quadratic TCFs for the numerically computed electrical and thermal conductivities at $T= 300$ K.

\section{Quantitative comparison of the transport coefficients}\label{comparesec}

\begin{figure}[htbp]
\includegraphics[trim={0cm 0cm 0cm  0cm},clip,width=9
cm]{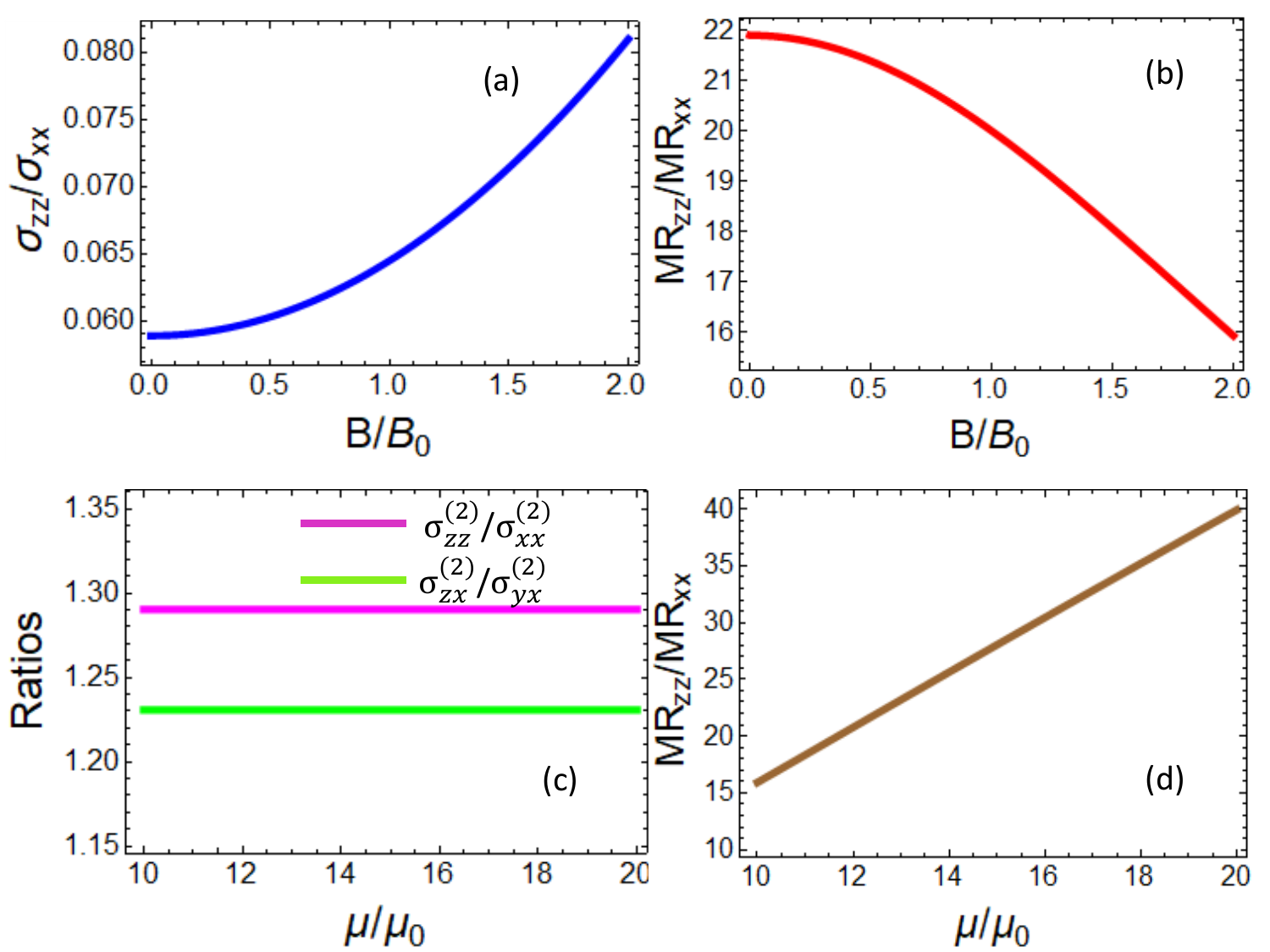}
\caption{ Plots of (a) the ratio of longitudinal magnetoconductivities ($\sigma_{zz}/\sigma_{xx}$) and  (b) ratio of longitudinal MR ($\text{MR}_{zz}/\text{MR}_{xx}$) of the respective orientations as a function of $B$ for ${\bf E} \ || \ {\bf B}$. (c) Ratio of quadratic-$B$ correction to longitudinal magnetoconductivities $(\sigma_{zz}^{(2)}/\sigma_{xx}^{(2)})$ and planar Hall conductivities ($\sigma_{zx}^{(2)}/\sigma_{yx}^{(2)}$) for the two orientation are independent of Fermi energy. (d) The ratio of longitudinal MR ($\text{MR}_{zz}/\text{MR}_{xx}$) as a function of Fermi energy. } 
\label{comp}
\end{figure}
In this section, we compare the magnitudes of transport coefficients for the two orientations by plotting their ratios with respect to magnetic field and Fermi energy. In Fig. \ref{comp}(a), the ratio of longitudinal magnetoconductivities $\sigma_{zz}$ and $\sigma_{xx}$ of the respective orientations is plotted as a function of $B$ for ${\bf E} \ || \ {\bf B}$. As $B\to0$, the ratio refers to that of the Drude conductivities. The value of $\sigma_{zz}^{(0)}$ is $\sim 6\%$ of that  $\sigma_{xx}^{(0)}$ for the chosen set of parameters, which implies a highly reduced Drude conductivity along $z$ axis. The ratio $\sigma_{zz}/\sigma_{xx}$ is a slowly increasing function of $B$. The increase is only $\sim 2\%$ upto $B/B_0=2$. Figure \ref{comp}(b) shows that the ratio of the magnetoresistances is a decreasing function of $B$. At very low $B$, $\text{MR}_{zz}$ is nearly 22 times of $\text{MR}_{xx}$. The ratio drops to $\sim 16$ at $B/B_0=2$. Thus, the direction in which the energy scales linearly with momentum in quadratic TCFs favours higher magnitude of MR and lower value of longitudinal magnetoconductivity. The measurements of MR and resistivity in presence of small magnetic field may hence act as probes  for detecting different symmetry axes of the system. 

We also observe that the ratio of quadratic-$B$ correction to longitudinal magnetoconductivities $(\sigma_{zz}^{(2)}/\sigma_{xx}^{(2)}$) and planar Hall conductivities ($\sigma_{zx}^{(2)}/\sigma_{yx}^{(2)}$)  for the two orientations are independent of Fermi energy as shown in Fig. \ref{comp}(c). The ratio $(\sigma_{zz}^{(2)}/\sigma_{xx}^{(2)}$) is higher than $(\sigma_{zx}^{(2)}/\sigma_{yx}^{(2)}$) . The ratio of longitudinal MR ($\text{MR}_{zz}/\text{MR}_{xx}$) is almost a linearly increasing function of Fermi energy as depicted in Fig.  \ref{comp}(d).

The ratio of the thermal coefficients show similar behaviour as that of the electrical ones since the longitudinal thermal and planar thermal Hall conductivities differ from their electrical counterparts simply by a factor ($L_0 T$) within the Sommerfeld approximation.

\section{Conclusions}\label{VII}
Triple-component semimetals host crossings of three energy bands near which the quasiparticles behave like pseudospin-1 fermions. In this work, we have dealt with the low-energy model for generalized triple-component fermions and obtained the Berry curvature, density of states and Fermi energy as function of carrier density. The DOS of dispersive bands varies with the Fermi energy as $D_{n}(\mu)\sim{\mu}^{2/n}$, where $n=1,2,3$. The Fermi energy as a function of electron density scales as $\mu\sim{n_e}^{n/n+2}$. These exponents are independent of the coefficients $\nu_z$ and $\alpha_n$.

We have specifically considered quadratic TCFs i.e., $n=2$ case and studied its Berry curvature induced electrical and thermal magnetotransport properties using semiclassical Boltzmann transport formalism. On the account of anisotropic energy spectrum of quadratic TCFs, we have obtained the transport coefficients for two different orientations of magnetic field -- ${\bf B}$ applied in $x$-$y$ and $x$-$z$ plane. Planar Hall effect is observed in both the orientations due to Berry curvature-modified magnetotransport. The PHC shows $B^2$ dependence in the lowest order in $B$ and an angular dependence of $\sin\beta\cos\beta$ as observed in other topological semimetals, where $\beta$ is the angle between applied electric and magnetic fields. 
The longitudinal magnetoconductivity also follows $B^2$ dependence due to Berry curvature, resulting in negative longitudinal MR. The longitudinal magnetoresistances corresponding to their respective $B$-field configurations ($\text{MR}_{xx}(\beta)$ and $\text{MR}_{zz}(\beta)$) vary as $\cos^2\beta$. We find that the out-of-plane magnetoconductivity shows an oscillatory dependence with $\beta$ for the second orientation of magnetic field, but it is  angle-independent for the first one. 

For the case of $B$ applied in $x$-$y$ plane, negative MR has the maximum value of about $-1.8\%$ whereas for $B$  confined in $x$-$z$ plane, the maximum MR reaches about $-29\%$. Hence it is evident that Berry curvature effects are considerably large for the latter case. Thus, the direction in which the energy scales linearly with momentum in quadratic TCFs favours higher magnitude of longitudinal MR. 

The Drude conductivities ${\sigma}_{xx}^{(0)}$ and ${\sigma}_{zz}^{(0)}$ are increasing functions of Fermi energy. The second-order corrections in longitudinal and planar magnetoconductivities and the absolute MRs for both the orientations decrease with the Fermi energy. This is attributed to the fact that Berry curvature effects decrease as the Fermi energy moves away from the band-touching nodes. The ratio ($\text{MR}_{zz}/\text{MR}_{xx}$) is almost a linearly increasing function of Fermi energy. A noteworthy feature of quadratic triple-component fermions which is typically absent in conventional systems is that certain transport coefficients and their ratios are independent of Fermi energy within the low-energy model. We observed that when $B$ is applied in the $x$-$z$ plane, quadratic $B$-correction in both $\sigma_{xx}$ and $\sigma_{yy}$ for ${\bf B}\ ||\ \hat{\bf z}$ is independent of Fermi energy. Furthermore, the ratio of quadratic-$B$ correction to longitudinal magnetoconductivities $(\sigma_{zz}^{(2)}/\sigma_{xx}^{(2)}$) and planar Hall conductivities ($\sigma_{zx}^{(2)}/\sigma_{yx}^{(2)}$)  corresponding to their respective orientations are also independent of Fermi energy. 

We also studied the longitudinal magneto-thermal conductivity and planar thermal  Hall conductivity in quadratic TCFs for both the magnetic field configurations. We note that LMTC and PTHC  show similar angular dependences as those of LMC and PHC and follow quadratic-$B$ dependence. Additionally, we find that Wiedemann-Franz law is valid for the numerically computed electrical and thermal conductivities at $T= 300$ K.
 
It is to be noted that in inversion symmetry broken and TRS preserved TCFs, the band touching nodes should appear in multiple of 4 just like the case of inversion symmetry broken Weyl semimetals. So, the magnitudes of the magnetotransport coefficients will be enhanced due to contribution from a greater number of nodes. Also, in such systems the triple-component nodes may not be at the same energy. This may result in different values of density of states, velocities and Berry curvatures at the Fermi surface of different nodes, unlike the TRS broken case. Moreover, in such systems, it is possible that the Fermi surface may intersect conduction band near one node (electron-like) and valence-band near the other (hole-like). So, we have to calculate the total magnetotransport coefficients considering low-energy Hamiltonians around each node separately.
\begin{center}
	{\bf ACKNOWLEDGEMENTS}
\end{center}
We would like to thank Sonu Verma for
useful discussions.

\begin{widetext}
\appendix{}

\section{Lorentz-force contribution}{\label{A}}
The components of $\mathbf{\Gamma}_{m}$ in Eq. (\ref{NDF}) for electrical and thermal transport are given below\cite{PHE1,PTHE3}. For brevity, we have dropped the band index $m$.\\
(a). For the planar Hall setup: ${\mathbf{E}}=E{\hat{x}}$, ${\mathbf{B}}= (B\cos\beta, B\sin\beta, 0)$ and $\boldsymbol{\nabla}{T}=0$,  
\begin{equation}
{\Gamma}_{z}=\frac{N({M}_{1}M_{2}+M_{3}M_{4})}{{(\frac{D}{\tau})^{2}}-{({\frac{eB\cos\beta}{m_{yz}}}-{\frac{eB\sin\beta}{m_{xz}}})}^{2}-{\frac{eB}{m_{zz}}}M_{4}},
\end{equation}\\
\begin{equation}
{\Gamma}_{y}=-\frac{\cos\beta[N{M_3}+\Gamma_{z}\frac{eB}{m_{zz}}]}{M_2},\hspace{1 cm}{\Gamma}_{x}=-\tan\beta{\Gamma}_{y},
\end{equation}
where $m_{ij}=\frac{1}{\hbar^{2}}\frac{\partial^{2}\epsilon_{\mathbf{k}}}{\partial{k_i}\partial{k_j}}$ is band-mass tensor, $N=\frac{e^{2}EB\tau}{D}$, ${M_1}=-\frac{\sin\beta}{m_{xx}}+\frac{\cos\beta}{m_{xy}}+\frac{eB\cos\beta}{\hbar}({C_2}\cos\beta-{C_1}\sin\beta)$,  ${M_2}=\frac{D}{\tau}-\frac{eB\sin\beta}{m_{xz}}+\frac{eB\cos\beta}{m_{yz}}$,  ${M_3}=\frac{eB\cos\beta}{\hbar}{C_3}+\frac{1}{m_{xz}}$,  ${M_4}=\frac{eB\sin2\beta}{m_{xy}}-\frac{eB{\cos}^{2}\beta}{m_{yy}}-\frac{eB{\sin}^{2}\beta}{m_{xx}}$, ${C_1}=\frac{\Omega_{x}}{m_{xx}}+\frac{\Omega_{y}}{m_{xy}}+\frac{\Omega_{z}}{m_{xz}}$, ${C_2}=\frac{\Omega_{x}}{m_{xy}}+\frac{\Omega_{y}}{m_{yy}}+\frac{\Omega_{z}}{m_{yz}}$ and ${C_3}=\frac{\Omega_{x}}{m_{xz}}+\frac{\Omega_{y}}{m_{yz}}+\frac{\Omega_{z}}{m_{zz}}$.\\
\\

(b). For the planar Hall setup: ${\mathbf{B}}= (B\cos\beta, B\sin\beta, 0)$, $\boldsymbol{\nabla}{T}={\nabla}{T}{\hat{x}}$ and ${\mathbf{E}}=0$,
\begin{equation}
{\Gamma}_{z}=\frac{N_{0}({\alpha}_{1}{\alpha}_{2}-{\alpha}_{3}{\alpha}_{4})}{{(\frac{D}{\tau})^{2}}-{({\frac{eB\cos\beta}{m_{yz}}}-{\frac{eB\sin\beta}{m_{xz}}})}^{2}-{\frac{eB}{m_{zz}}}{\alpha}_{4}},
\end{equation}\\
\begin{equation}
{\Gamma}_{y}=\frac{\cos\beta[{N}_{0}(\frac{1}{m_{xz}}+\frac{e B {C_3}\cos\beta}{\hbar})-{\Gamma}_{z}\frac{eB}{m_{zz}}]}{({\frac{eB\cos\beta}{m_{yz}}}-\frac{D}{\tau}-{\frac{eB\sin\beta}{m_{xz}}})},
\end{equation}
\begin{equation}
{\Gamma}_{x}=-\tan\beta{\Gamma}_{y},
\end{equation}
where $N_{0}=eB\tau{\nabla}{T}\frac{(\epsilon_{k}-\mu)}{DT}$, ${\alpha_1}=\frac{\sin\beta}{m_{xx}}-\frac{\cos\beta}{m_{xy}}+\frac{eB\cos\beta}{\hbar}(-{C_2}\cos\beta+{C_1}\sin\beta)$,  ${\alpha_2}=-\frac{D}{\tau}+\frac{eB\cos\beta}{m_{yz}}-\frac{eB\sin\beta}{m_{xz}}$,  ${\alpha_3}=\frac{eB\cos\beta}{\hbar}{C_3}+\frac{1}{m_{xz}}$,  ${\alpha_4}=\frac{eB\sin2\beta}{m_{xy}}-\frac{eB{\cos}^{2}\beta}{m_{yy}}-\frac{eB{\sin}^{2}\beta}{m_{xx}}$.\\
\\
\end{widetext}

\end{document}